\documentclass[fleqn,usenatbib]{mnras}
\usepackage{newtxtext,newtxmath}

\usepackage[T1]{fontenc}
\usepackage{ae,aecompl}
\usepackage{multirow}

\usepackage{graphicx}	
\usepackage{amsmath}	
\usepackage{amssymb}	
\usepackage{ulem} 

\usepackage[switch,columnwise]{lineno}

\newcommand{\CIV}{C\,{\small IV}\,$\lambda$1549}

\newcommand{\MgII}{Mg\,{\small II}\,$\lambda$2800}

\newcommand{\hbeta}{H{$\beta$}}

\title[Candidate Periodic Quasars from DES+SDSS]{Candidate Periodically Variable Quasars from the Dark Energy Survey and the Sloan Digital Sky Survey}

\author[Chen et al.]{
\parbox{\textwidth}{
\Large
Yu-Ching~Chen,$^{1,2}$\thanks{E-mail: ycchen@illinois.edu (YCC), xinliuxl@illinois.edu (XL)}
Xin~Liu,$^{1,2}$
Wei-Ting~Liao,$^{1}$
A.~Miguel~Holgado,$^{1,2}$
Hengxiao~Guo,$^{1,2}$
Robert~A.~Gruendl,$^{1,2}$
Eric~Morganson,$^{1,2}$
Yue~Shen,$^{1,2}$\thanks{Alfred P. Sloan Research Fellow}
Kaiwen~Zhang,$^{3}$
Tim~M.~C.~Abbott,$^{4}$
Michel~Aguena,$^{5,6}$
Sahar~Allam,$^{7}$
Santiago~Avila,$^{8}$
Emmanuel~Bertin,$^{9,10}$
Sunayana~Bhargava,$^{11}$
David~Brooks,$^{12}$
David~L.~Burke,$^{13,14}$
Aurelio~Carnero~Rosell,$^{15}$
Daniela~Carollo,$^{16}$
Matias~Carrasco~Kind,$^{1,2}$
Jorge~Carretero,$^{17}$
Matteo~Costanzi,$^{18,19}$
Luiz~N.~da Costa,$^{6,20}$
Tamara~M.~Davis,$^{21}$
Juan~De~Vicente,$^{15}$
Shantanu~Desai,$^{22}$
H.~Thomas~Diehl,$^{7}$
Peter~Doel,$^{12}$
Spencer~Everett,$^{23}$
Brenna~Flaugher,$^{7}$
Douglas~Friedel,$^{2}$
Joshua~Frieman,$^{7,24}$
Juan~Garc\'ia-Bellido,$^{8}$
Enrique~Gaztanaga,$^{25,26}$
Karl~Glazebrook,$^{27}$
Daniel~Gruen,$^{28,13,14}$
Gaston~Gutierrez,$^{7}$
Samuel~R.~Hinton,$^{21}$
Devon~L.~Hollowood,$^{23}$
David~J.~James,$^{29}$
Alex~G.~Kim,$^{30}$
Kyler~Kuehn,$^{31,32}$
Nikolay~Kuropatkin,$^{7}$
Geraint~F.~Lewis,$^{33}$
Christopher~Lidman,$^{34}$
Marcos~Lima,$^{5,6}$
Marcio~A.~G.~Maia,$^{6,20}$
Marisa~March,$^{35}$
Jennifer~L.~Marshall,$^{36}$
Felipe~Menanteau,$^{1,2}$
Ramon~Miquel,$^{37,17}$
Antonella~Palmese,$^{7,24}$
Francisco~Paz-Chinch\'{o}n,$^{38,2}$
Andr\'es~A.~Plazas,$^{39}$
Eusebio~Sanchez,$^{15}$
Michael~Schubnell,$^{40}$
Santiago~Serrano,$^{25,26}$
Ignacio~Sevilla-Noarbe,$^{15}$
Mathew~Smith,$^{41}$
Eric~Suchyta,$^{42}$
Molly~E.~C.~Swanson,$^{2}$
Gregory~Tarle,$^{40}$
Brad~E.~Tucker,$^{34}$
Tamas~Norbert~Varga,$^{43,44}$
and Alistair~R.~Walker$^{4}$
\begin{center} (DES Collaboration) \end{center}
}
\vspace{0.4cm}
\\
\parbox{\textwidth}{
\scriptsize
$^{1}$ Department of Astronomy, University of Illinois at Urbana-Champaign, 1002 W. Green Street, Urbana, IL 61801, USA\\
$^{2}$ National Center for Supercomputing Applications, 1205 West Clark St., Urbana, IL 61801, USA\\
$^{3}$ Department of Physics, University of Illinois at Urbana-Champaign, 1110 W. Green Street, Urbana, IL 61801, USA\\
$^{4}$ Cerro Tololo Inter-American Observatory/NSF’s NOIRLab, Casilla 603, La Serena, Chile\\
$^{5}$ Departamento de F\'isica Matem\'atica, Instituto de F\'isica, Universidade de S\~ao Paulo, CP 66318, S\~ao Paulo, SP, 05314-970, Brazil\\
$^{6}$ Laborat\'orio Interinstitucional de e-Astronomia - LIneA, Rua Gal. Jos\'e Cristino 77, Rio de Janeiro, RJ - 20921-400, Brazil\\
$^{7}$ Fermi National Accelerator Laboratory, P. O. Box 500, Batavia, IL 60510, USA\\
$^{8}$ Instituto de Fisica Teorica UAM/CSIC, Universidad Autonoma de Madrid, 28049 Madrid, Spain\\
$^{9}$ CNRS, UMR 7095, Institut d'Astrophysique de Paris, F-75014, Paris, France\\
$^{10}$ Sorbonne Universit\'es, UPMC Univ Paris 06, UMR 7095, Institut d'Astrophysique de Paris, F-75014, Paris, France\\
$^{11}$ Department of Physics and Astronomy, Pevensey Building, University of Sussex, Brighton, BN1 9QH, UK\\
$^{12}$ Department of Physics \& Astronomy, University College London, Gower Street, London, WC1E 6BT, UK\\
$^{13}$ Kavli Institute for Particle Astrophysics \& Cosmology, P. O. Box 2450, Stanford University, Stanford, CA 94305, USA\\
$^{14}$ SLAC National Accelerator Laboratory, Menlo Park, CA 94025, USA\\
$^{15}$ Centro de Investigaciones Energ\'eticas, Medioambientales y Tecnol\'ogicas (CIEMAT), Madrid, Spain\\
$^{16}$ INAF, Astrophysical Observatory of Turin, I-10025 Pino Torinese, Italy\\
$^{17}$ Institut de F\'{\i}sica d'Altes Energies (IFAE), The Barcelona Institute of Science and Technology, Campus UAB, 08193 Bellaterra (Barcelona) Spain\\
$^{18}$ INAF-Osservatorio Astronomico di Trieste, via G. B. Tiepolo 11, I-34143 Trieste, Italy\\
$^{19}$ Institute for Fundamental Physics of the Universe, Via Beirut 2, 34014 Trieste, Italy\\
$^{20}$ Observat\'orio Nacional, Rua Gal. Jos\'e Cristino 77, Rio de Janeiro, RJ - 20921-400, Brazil\\
$^{21}$ School of Mathematics and Physics, University of Queensland,  Brisbane, QLD 4072, Australia\\
$^{22}$ Department of Physics, IIT Hyderabad, Kandi, Telangana 502285, India\\
$^{23}$ Santa Cruz Institute for Particle Physics, Santa Cruz, CA 95064, USA\\
$^{24}$ Kavli Institute for Cosmological Physics, University of Chicago, Chicago, IL 60637, USA\\
$^{25}$ Institut d'Estudis Espacials de Catalunya (IEEC), 08034 Barcelona, Spain\\
$^{26}$ Institute of Space Sciences (ICE, CSIC),  Campus UAB, Carrer de Can Magrans, s/n,  08193 Barcelona, Spain\\
$^{27}$ Centre for Astrophysics \& Supercomputing, Swinburne University of Technology, Victoria 3122, Australia\\
$^{28}$ Department of Physics, Stanford University, 382 Via Pueblo Mall, Stanford, CA 94305, USA\\
$^{29}$ Center for Astrophysics $\vert$ Harvard \& Smithsonian, 60 Garden Street, Cambridge, MA 02138, USA\\
$^{30}$ Lawrence Berkeley National Laboratory, 1 Cyclotron Road, Berkeley, CA 94720, USA\\
$^{31}$ Australian Astronomical Optics, Macquarie University, North Ryde, NSW 2113, Australia\\
$^{32}$ Lowell Observatory, 1400 Mars Hill Rd, Flagstaff, AZ 86001, USA\\
$^{33}$ Sydney Institute for Astronomy, School of Physics, A28, The University of Sydney, NSW 2006, Australia\\
$^{34}$ The Research School of Astronomy and Astrophysics, Australian National University, ACT 2601, Australia\\
$^{35}$ Department of Physics and Astronomy, University of Pennsylvania, Philadelphia, PA 19104, USA\\
$^{36}$ George P. and Cynthia Woods Mitchell Institute for Fundamental Physics and Astronomy, and Department of Physics and Astronomy, Texas A\&M University, College Station, TX 77843,  USA\\
$^{37}$ Instituci\'o Catalana de Recerca i Estudis Avan\c{c}ats, E-08010 Barcelona, Spain\\
$^{38}$ Institute of Astronomy, University of Cambridge, Madingley Road, Cambridge CB3 0HA, UK\\
$^{39}$ Department of Astrophysical Sciences, Princeton University, Peyton Hall, Princeton, NJ 08544, USA\\
$^{40}$ Department of Physics, University of Michigan, Ann Arbor, MI 48109, USA\\
$^{41}$ School of Physics and Astronomy, University of Southampton,  Southampton, SO17 1BJ, UK\\
$^{42}$ Computer Science and Mathematics Division, Oak Ridge National Laboratory, Oak Ridge, TN 37831\\
$^{43}$ Max Planck Institute for Extraterrestrial Physics, Giessenbachstrasse, 85748 Garching, Germany\\
$^{44}$ Universit\"ats-Sternwarte, Fakult\"at f\"ur Physik, Ludwig-Maximilians Universit\"at M\"unchen, Scheinerstr. 1, 81679 M\"unchen, Germany\\
}
}

\date{Accepted XXX. Received YYY; in original form ZZZ}

\pubyear{2020}

\usepackage{eso-pic}

\AddToShipoutPictureBG*{%
  \AtPageUpperLeft{%
    \hspace{0.75\paperwidth}%
    \raisebox{-3.5\baselineskip}{%
      \makebox[0pt][l]{\textnormal{DES 2020-0544}}
}}}%

\AddToShipoutPictureBG*{%
  \AtPageUpperLeft{%
    \hspace{0.75\paperwidth}%
    \raisebox{-4.5\baselineskip}{%
      \makebox[0pt][l]{\textnormal{FERMILAB-PUB-20-179-AE}}
}}}%

\begin{document}
\label{firstpage}
\pagerange{\pageref{firstpage}--\pageref{lastpage}}
\maketitle

\begin{abstract}
Periodically variable quasars have been suggested as close binary supermassive black holes. We present a systematic search for periodic light curves in 625 spectroscopically confirmed quasars with a median redshift of 1.8 in a 4.6 deg$^2$ overlapping region of the Dark Energy Survey Supernova (DES-SN) fields and the Sloan Digital Sky Survey Stripe 82 (SDSS-S82). Our sample has a unique 20-year long multi-color ($griz$) light curve enabled by combining DES-SN Y6 observations with archival SDSS-S82 data. The deep imaging allows us to search for periodic light curves in less luminous quasars (down to $r{\sim}$23.5 mag) powered by less massive black holes (with masses $\gtrsim10^{8.5}M_{\odot}$) at high redshift for the first time. We find five candidates with significant (at $>$99.74\% single-frequency significance in at least two bands with a global p-value of $\sim$7$\times10^{-4}$--3$\times10^{-3}$ accounting for the look-elsewhere effect) periodicity with observed periods of $\sim$3--5 years (i.e., 1--2 years in rest frame) having $\sim$4--6 cycles spanned by the observations. If all five candidates are periodically variable quasars, this translates into a detection rate of ${\sim}0.8^{+0.5}_{-0.3}$\% or ${\sim}1.1^{+0.7}_{-0.5}$ quasar per deg$^2$. Our detection rate is 4--80 times larger than those found by previous searches using shallower surveys over larger areas. This discrepancy is likely caused by differences in the quasar populations probed and the survey data qualities. We discuss implications on the future direct detection of low-frequency gravitational waves. Continued photometric monitoring will further assess the robustness and characteristics of these candidate periodic quasars to determine their physical origins.
\end{abstract}




\section{Introduction}

Supermassive black holes (SMBHs) with masses ${\sim}10^5$--$10^9M_{\odot}$ are commonly found at the hearts of massive galaxies \citep{kormendy95}. When galaxies merge, their central black holes should collide and form binaries\citep{begelman80}. These binary supermassive black holes (BSBHs) are particularly interesting, because they create distortions in spacetime, known as gravitational waves  \citep[GWs][]{Einstein1916,Einstein1918}, that have the highest strain amplitude and make the loudest GW sirens in the universe. More massive binaries are pulsar-timing array (PTA) sources \citep[e.g.,][]{Arzoumanian2018a}. Less massive binaries are targeted by space-based experiments such as LISA \citep{Klein2016}. They provide a laboratory to directly test strong-field general relativity \citep{hughes09,Centrella2010}. 

BSBHs are also important for multiple topics in cosmology and galaxy formation \citep{colpi09}. First, since BSBHs are expected from galaxy mergers, their abundance provides an important check for the LCDM hierarchical structure formation paradigm \citep{Volonteri2009,yu11}. Second, BSBHs are believed to have a significant dynamical impact on the stellar structures of galactic nuclei. They scour out stellar cores as they eject stars through three body interactions \citep{Kormendy2009}. Third, because gas-rich mergers are expected to trigger strong gas inflows to the galactic centers, BSBHs offer a unique laboratory to study merger-induced accretion and black hole growth and their possible effects on the evolution of their host galaxies (e.g., through strong outflows or so-called ``feedback''). Finally, the successive dynamical evolution of BSBHs in galaxy mergers is also of great interest \citep{DEGN}. 

While the formation of BSBHs is inevitable, direct evidence has so far been elusive. No confirmed case is known in the GW-dominated regime (a binary is so close that the orbital decay is driven by GW emission). A critical unsolved problem is that the orbital decay of a BSBH may significantly slow down or even stall at $\sim$parsec scales, i.e., the so-called ``final-parsec'' problem \citep{begelman80,milosavljevic01,yu02}. When a binary has run out of stars to interact with but has not approached close enough to significantly emit significant gravitational radiation, there is no obvious method for its orbit to decay. This long-standing conundrum poses the largest uncertainty on the abundance of BSBH mergers as GW sources. In theory, the bottleneck may be overcome in gaseous environments \citep[e.g.,][]{gould00,Cuadra2009,Chapon2013,delValle2015}, in triaxial or axisymmetric galaxies \citep[e.g.,][]{Khan2016,Kelley2017}, and/or by interacting with a third BH in hierarchical mergers \citep[e.g.,][]{blaes02,Kulkarni2012,Bonetti2018}. Observations of BSBHs are needed to test these theories and to verify their feasibility and efficiencies in solving the final-parsec problem.

Quantifying the occurrence rate of BSBHs is important for understanding the various gas and stellar dynamical processes to solve the final-parsec problem. However, the physical separations of BSBHs that are gravitationally bound to each other ($\lesssim$ a few pc) are too small for direct imaging. Even with resolution of 10 microarcseconds cannot resolve BSBHs except for in the local universe \citep{burke11}. CSO 0402+379 (discovered by VLBI as a double flat-spectrum radio source separated by 7 pc) remains the only secure case known \citep{rodriguez06,Bansal2017}. While great strides have been made in identifying dual active galactic nuclei (AGN) -- progenitors of BSBHs at $\gtrsim$kpc scales \citep[e.g.,][]{komossa03,ballo04,hudson06,Liu2010a,Liu2011,Liu2013,Comerford2015,Fu2015a,Muller-Sanchez2015,Koss2016,Ellison2017,Liu2018b,Silverman2020}, there is no confirmed BSBH at milli-pc scales, i.e., in the GW regime \citep[e.g.,][]{Bogdanovic2015,Komossa2016}. 

Periodic quasar light curves have long been proposed as candidate milli-pc BSBHs. Periodicity may arise from accretion rate changes \citep[e.g.,][]{MacFadyen2008,Shi2012,Roedig2012,DOrazio2013,Farris2014,Tang2018}, and/or relativistic Doppler boost from the highly relativistic motion of gas in the mini accretion disk around the smaller BH in a binary \citep[e.g.,][]{DOrazio2015a}. While $\sim$150 periodic quasars have been found as BSBH candidates \citep[e.g.,][]{valtonen08,Graham2015,Graham2015a,Liutt2015,Liutt2016,Charisi2016,Zheng2016,Li2019,Liutt2019}, most of the known candidates have been shown to be subject to false positives due to stochastic quasar variability \citep[e.g.,][]{Vaughan2016,Barth2018,Goyal2018,Liutt2018}. Furthermore, previous surveys were only sensitive to the most massive quasars at high redshift ($z\gtrsim2$) which should have already gone through their major merger process \citep[e.g.,][]{shen09}. The physical origin of the candidate periodicity has also been largely uncertain \citep[e.g.,][]{Graham2015,Charisi2018}.

In this paper, we present a systematic search for periodically variable quasars as candidates for milli-pc BSBHs in the GW-dominated regime. We combine the newly obtained, highly sensitive imaging from the Dark Energy Survey Supernova \citep[DES-SN,][]{Kessler2015} project with archival Sloan Digital Sky survey Stripe 82 (SDSS-S82) data. 
The deep Dark Energy Survey(DES) imaging allows us to search for periodic light curves in less luminous quasars (down to $r{\sim}$23.5 mag) which are powered by the less massive and more common SMBH populations at high redshift for the first time. 

Compared to previous studies, our candidates are expected to be more robust, because DES has higher sensitivities (generally $>$2 mag deeper), and when combined with archival Sloan Digital Sky Survey(SDSS) data, a 2--5 times longer time baselines than previous shallower surveys of larger area. \autoref{tab:compare} compares our work against previous studies. The combination of DES and SDSS represents the best dataset among all currently available synoptic surveys in terms of time baseline and sensitivity. High data quality is crucial both for rejecting false positives and for recovering false negatives to minimize the systematic error of detections.

\begin{table*}
\begin{center}
 \resizebox{1.0\textwidth}{!}{%
    \begin{tabular}{ | p{2cm} | p{1.1cm} | p{2.05cm} | p{3.7cm} | p{1.9cm} | p{1.4cm} |p{0.9cm} | p{0.9cm} |}
    \hline\hline
Program & Time baseline & Telescope\& Aperture & Single-epoch 5$\sigma$ Point-source Depth & Cadence & Mean Cadence & Area (deg$^2$) & Band  \\ \hline
CRTS$^{{\rm [1],[2],[3]}}$ & 9 years & MLS/CSS/SSS, 1.5m/0.7m/0.5m & $\sim$20 (Vega) & 7 days & 13 days & 33,000 & V \\ \hline
PTF$^{{\rm [4]}}$   & 3.8 years & SOS, 1.2m & 21.3, 20.6 (Vega) & 5 days & 3--50 days & 2,700 & gR \\ \hline
PanSTARRS1 MD09$^{{\rm [5],[6],[7]}}$ & 4.2 years & Haleakala, 1.8m & 22.0, 21.8, 21.5, 20.9, 19.7 (AB) & 3 days & 6 days & 8 &  grizY \\ \hline
This Work (DES+SDSS) & 20 years & Blanco/APO, 4m/2.5m & 24.3, 24.1, 23.5, 22.9 (AB) for DES / 22.2, 22.2, 21.3, 20.5 (AB) for SDSS &  7 days for DES / 4 days for SDSS & 35 days for DES+SDSS & 4.6 & griz 
\\
\hline
    \end{tabular}
    }
    \caption{This work in comparison to previous studies. [1]: \citet{Graham2015a}, [2]: \citet{Graham2015}, [3]: \citet{Zheng2016}; [4]: \citet{Charisi2016}; [5]: \citet{Liutt2015}, [6]: \citet{Liutt2016}, [7]: \citet{Liutt2019}. While \citet{Liutt2019} have presented follow-up observations that have extended the time baseline for a few initial candidates, quoted here are values appropriate for the full parent sample. Cadence is the stated rate over time baseline without seasonal gaps. Mean cadence is the average rate over total time baseline including seasonal gaps.}
    \label{tab:compare}
    
\end{center}
\end{table*}

The paper is organized as follows. \S \ref{sec:obs} describes our observations, sample selection, and basic properties. We discuss our data analysis method in \S \ref{sec:method}. We then present our results in \S \ref{sec:result}, followed by light curve modeling in  \S \ref{sec:modeling} and discussions of their implications in \S \ref{sec:discussion}. Finally, we summarize the main results and suggest directions for future work in \S \ref{sec:sum}. A concordance $\Lambda$CDM cosmology with $\Omega_m = 0.3$, $\Omega_{\Lambda} = 0.7$, and $H_{0}=70$ km s$^{-1}$ Mpc$^{-1}$ is assumed throughout. We use the AB magnitude system \citep{Oke1974} unless otherwise noted.

\section{Observations, Sample Selection, and Sample Properties}\label{sec:obs}

\subsection{Program Design: Combining the Dark Energy Survey with SDSS}\label{subsec:design}

We combine the new DES-SN Year 6 (Y6) multi-color ($griz$) imaging (spanning 2012--2019) with archival SDSS-S82 data (spanning 1998--2007). There is a 5-yr gap between SDSS-S82 and DES-SN. This is being partially filled by publicly available archival data (from CRTS (2005-2013), PTF (2009--2012), PS1 3$\pi$ (2009--2014), and/or ZTF (2017--present) for the brighter quasars in the sample. The total time baseline extends $\sim$20 yr. 

BSBHs with (total) masses of $\sim10^8$--$10^9M_{\odot}$ at redshift $z{\gtrsim}1$ are generally expected around the time of pre-decoupling \citep[e.g.,][]{Kocsis2011a}, i.e., when the GW inspiral time $t_{{\rm gw}}{>}t_{{\rm visc}}$, where $t_{{\rm visc}}$ is the viscous timescale of the accretion disk. For a typical quasar at $z{\sim}$1, the baseline spans ${\sim}10$ yr (rest-frame) to enclose ${\gtrsim}$5 cycles for a period of $\lesssim$2 yr. A coverage of 3--5 cycles is generally recommended to minimize false periodicity due to stochastic red-noise variability \citep{Vaughan2016}.

\subsubsection{Dark Energy Survey}
\label{sec:DES}

DES is a six-year (2013--2019, not counting Science Verification in 2012) optical imaging survey of the Southern Hemisphere \citep{flaugher2005,Diehl2019}. Using the Dark Energy Camera \citep[DECam,][]{Flaugher2015} on the 4-m Blanco telescope at Cerro Tololo Inter-American observatory, DES studies the properties of dark energy via different probes such as type Ia supernovae, weak lensing, galaxy clusters, and baryon acoustic oscillation. Images taken with DES have been transferred to the National Center for Supercomputing Applications for processing and release \citep{morganson2018}. 
The first DES public data release \citep[DES DR1,][]{DES_DR1} contains single-epoch image, co-added images, co-added source catalogs, and associated products collected over the first three years of DES operations. The second DES public data release (DES DR2) is scheduled for 2021.

DES consists of two surveys -- a wide-field survey of 5000 deg$^2$ in \textit{grizY} bands and a time-domain, also called SN field, survey of $\sim$30 deg$^2$ with high cadences ($\sim$7 days) in the \textit{griz} bands. The typical single-epoch 5-$\sigma$ depth for the wide field is 24.3, 24.1, 23.5, 22.9 mag in the \textit{griz} bands \citep{DES_DR1}. Sources in each exposure have been calibrated using a forward modeling technique \citep{DES_photo_cal} and placed on the AB system \citep{AB_system}. The single-epoch photometric statistical precision is 7.3, 6.1, 5.9, 7.3 mmag in the \textit{griz} bands \citep{DES_DR1}. 

Among the ten DES-SN fields in the time-domain survey, two (S1 and S2) are within the SDSS-S82 footprint. This 4.6 deg$^2$ overlapping region between DES-SN and SDSS-S82 provides the long time baseline needed for a high-fidelity search of periodic light curves against a background of red noise stochastic quasar variability.

\subsubsection{Sloan Digital Sky Survey}
\label{sec:SDSS} 

The SDSS-S82 is an area of $\sim$300 deg$^2$ on the Celestial Equator. The SDSS imaging survey on Stripe 82 region were conducted by the 2.5-m telescope at the Apache Point Observatory from September 1998 to September 2007 with total epochs of $\sim$70--90 in the \textit{ugriz} bands \citep{SDSS_DR5,SDSS_S82_starcatalog,SDSS_SN_tech}. The typical single-epoch 5-$\sigma$ depth is 22.0, 22.2, 22.2, 21.3, 20.5 mag in the \textit{ugriz} bands
\citep{SDSS_DR5}. All SDSS magnitudes have been calibrated to the AB magnitude system \citep{SDSS_DR7} with photometric accuracy of $\sim$0.02--0.03 mag \citep{SDSS_S82_starcatalog}.

\subsubsection{Supplementary Data from CRTS, PTF, PS1, and/or ZTF}

We adopt archival data from other imaging surveys. These include the Catalina Real Time Transient Survey \citep[CRTS;][]{Drake2009}, the Palomar Transient Factory \citep[PTF;][]{PTF}, Panoramic Survey Telescope and Rapid Response System  \citep[PS1;][]{PS1_survey,PS1_data}, and the Zwicky Transient Facility \citep[ZTF;][]{ZTF_DR1}. They fill the cadence gap between SDSS and DES and serve as independent double checks. We do not include them in our baseline analysis, however, in view of systematic uncertainties due to telescope system conversions (see \S \ref{sec:mag_convert}).

The CRTS covers 500 million objects with V-band magnitudes between 11.5 and 21.5 in an area of 33,000 deg$^2$. The second CRTS public data release contains data from the seven years of photometry (2005$-$2013).  The PTF observed between 2013 and 2015 with the Samuel Oschin 48-inch telescope in 2 filters (\textit{gR}) at Palomar Observatory. The 3-$\pi$ Steradian Survey in PS1 covers the sky north of $-30$ deg with 60 total epochs in the \textit{grizy$_{P1}$} bands from 2009 to 2014. The mean 5-$\sigma$ depth of the single-epoch 3-$\pi$ Survey in \textit{grizy$_{P1}$} is 22.0, 21.8, 21.5, 20.9, and 19.7 mag, respectively \citep{PS1_survey}. ZTF uses the Samuel Oschin 48-inch telescope and a new camera with a 47 deg$^2$ field of view to scan the northern sky to median depths of $g\sim$20.8 mag and $r\sim$20.6 mag. The ZTF public data release 1 contains observations spanning March 2018 to December 2018. 




\subsection{Quasar Sample Selection and Sample Properties}

We start from the spectroscopically confirmed quasars compiled from various quasar catalogs. These include the SDSS DR7 and DR14 quasar catalogs \citep{SDSS_DR7Q,SDSS_DR14Q}, the OzDES DR1 \citep{OzDES_DR1}, and the Million quasar catalog \citep[v6.2, 22 May 2019,][]{HMQ}. 
We then cross-match the spectroscopically confirmed quasars with the DES coadd catalog (Y3A1\_COADD\_OBJECT\_SUMMARY) in the overlapping region of DES-SN S1 and S2, and SDSS-S82. 

To ensure the artifact-free DES sample, we require 
\begin{enumerate}
\item \texttt{FLAGS\_\{G,R,I,Z,Y\}} = 0, and 
\item \texttt{IMAFLAGS\_ISO\_\{G,R,I,Z,Y\}} = 0.
\end{enumerate}
Both parameters are produced by \textsc{sextractor} \citep{Bertin1996}. \texttt{FLAGS} is an internal flag produced during the source extraction. \texttt{IMAFLAGS\_ISO} is an external flag containing the values of flag map pixels that overlap the isophotal area of a given detected object. Both are used to reject saturated sources, blended sources, or sources affected by bad pixels. Our initial selection has resulted in 763 spectroscopically confirmed quasars in the overlapping region of DES-SN S1 and S2, and SDSS-S82.


\subsubsection{Light Curve Generation}

We generate the DES light curves using the single-epoch photometry between the Scientific Verification phase (SV) and Y6 from the Year 6 Annual release 1 (Y6A1). We adopt the following selection criteria:
\begin{enumerate}
\item \texttt{FLAGS} = 0, and 
\item \texttt{IMAGLAGS\_ISO} = 0.
\end{enumerate}
On average, 3.7 percent of data points are rejected by the two criteria listed above for each light curve. We use the Forward Global Calibration Method \citep{DES_photo_cal} for photometric calibration. We query the SDSS archival database using Butler's script\footnote{\url{http://butler.lab.asu.edu/qso_selection/index.html}} for downloading S82 light curves.

\begin{figure}
\centering
\includegraphics[width=0.5\textwidth]{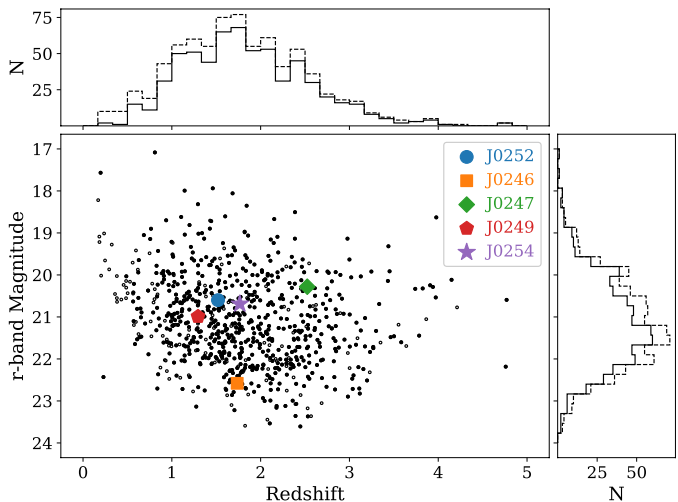}
\caption{Distribution of the redshift and magnitude for the final parent sample of 625 quasars (filled circles and solid lines) and the initial sample of 763 quasars (open circles and dashed lines). The  The five periodic quasar candidates, including J0252 reported in \citet{Liao2020}, are marked in colored symbols.
\label{fig:redshift_mag}}
\end{figure}

We require at least 50 DES epochs and 30 SDSS epochs in each of at least two bands. The median total numbers of imaging epochs in the final parent quasar sample are 80, 80, 79, 79 in the $griz$ bands from SDSS and 132, 138, 135, 140 in the $griz$ bands from DES.
Our final quasar sample contains 625 
quasars with sufficient epochs. Figure \ref{fig:redshift_mag} shows the redshift and magnitude distributions of the 625 quasars. 
We remove 5-$\sigma$ outliers from the running median in the light curves in each band and bin the different observations within the same Julian date to achieve a better S$/$N.


\subsubsection{Magnitude Conversion between Different Telescope Systems}\label{sec:mag_convert}

To stitch together the light curves from different telescope systems, we estimate the magnitude correction using their transmission curves. Since quasars may have different spectral energy distributions (SEDs), we estimate the magnitude correction based on the individual quasar spectra. We assume that temporal color variation is negligible.

Among the 625 quasars, 619 have optical spectra from the SDSS DR14. We convolve each quasar spectrum with the transmission curves of DES and SDSS to calculate the synthetic magnitudes. 
To minimize bias due to noise, we remove the data points with $and\_mask > 0$ and reject 3-$\sigma$ outliers using running median with a window size of 11 spectral pixels over the smoothed spectra. We then calculate the magnitude difference in each band between two systems to obtain the magnitude correction. The mean magnitude correction of our sample from SDSS to DES is $-$0.029, $-$0.034, $-$0.031, and $-$0.012 in the \textit{griz} bands. The 1-$\sigma$ statistical errors are also propagated to the SDSS measurements.

All light curves have been corrected to be on the DES system for consistency. For the six quasars without SDSS spectra, we assume the average magnitude correction factor from the other quasars with available spectra. 




\section{Method and Analysis}\label{sec:method}

\subsection{Generalized Lomb-Scargle Periodogram}

The Lomb-Scargle periodogram \citep{Lomb-Scargle} is widely used for periodicity detection in unevenly sampled data. It is equivalent to fitting sinusoidal waves in the form of $a\text{ sin }\omega t+b\text{ cos }\omega t$. We adopt the generalized Lomb-Scargle (GLS) periodogram \citep{Generalised_Lomb-Scargle} contained in the \textit{astroML} package \citep{astroML}. It provides more accurate frequency prediction. Compared to the classic method, the GLS periodogram takes an offset and weights ($\chi^2$ fit) into consideration. The searched frequency range of GLS periodogram is from 1/0.75 year$^{-1}$ to 1/T$_{\rm span}$ year$^{-1}$, where T$_{\rm span}$ is the time baseline, with uniform sampling steps equivalent to the number of observations.

To select periodic quasars, instead of making a flat cut on the normalized periodograms, we compare the power to those of the simulated light curves (see \S \ref{subsubsec:simulation}) accounting for quasars' stochastic red noise variability.  We verify our results using the multi-band Lomb-Scargle periodogram of \citet{VanderPlas2015} and the revised GLS periodogram method\footnote{\url{http://butler.lab.asu.edu/qso_period/}} (GLSdeDRW) adopted by \citet{Zheng2016}. The multi-band GLS yields consistent results with those from the single-band GLS analysis. The revised GLS approach verifies the periodic candidates but overestimates the significance of the periodicity because of the white noise assumption adopted in \citet{Zheng2016}.

\subsubsection{Simulating Quasar Light Curves with Tailored Variability Properties}\label{subsubsec:simulation}

Quasar variability can be modeled as a continuous time first-order auto-regressive process \cite[CAR(1),][]{Kelly2009} or so-called Damped Random Walk \cite[DRW,][]{MacLeod2010} model. It is a random-walk model added with a correction term pushing the variation back to the mean value. The CAR(1) model can be described by the following differential equation: 

\begin{equation}\label{eq:drw}
    dX(t) = -\frac{1}{\tau} X(t)dt + \sigma \sqrt{dt} \epsilon(t) +b dt,
\end{equation}
where $X(t)$ is the quasar flux, $\tau$ is the relaxation timescale, $\tau \sigma^2/2$ is the variance, $\tau b$ is the mean value, and $\epsilon(t)$ is a white noise process with a Gaussian distribution. 

We estimate the CAR(1) model parameters using a maximum likelihood method. Following \citet{Kelly2009}, we construct the likelihood function $p(x_1,...,x_i|b,\sigma,\tau)$ where $x_i$ is the quasar flux at epoch $i$. We estimate the characteristic timescale and variance directly in the time domain because the anomalous power due to uneven sampled seasonal gaps and cadences might bias the parameter fitting. We then employ the Bayesian approach to obtain the posterior distribution for $b$, $\sigma$, and $\tau$ using a Markov Chain Monte Carlo (MCMC) method with the \textit{emcee} package. We adopt a non-informative prior \citep{emcee}. We burn the first half chains and exclude the top 5\% and bottom 5\% parameters to avoid biases from initial position of walkers. We generate 50,000 simulated light curves by integrating Equation \ref{eq:drw} with the parameters $b$, $\sigma$, and $\tau$ drawn from the posterior distribution. The simulated light curves are then matched to the real observing cadence with measurement errors injected.

\subsubsection{Statistical Significance of Periodicity}

We estimate the statistical significance of any periodicity detected in the GLS periodograms using the simulated light curves with tailored variability properties. We calculate the significance value at each given frequency bin. We select candidates with $>$99.74\% significance in at least two bands. We reject any candidate periodicity detected at $<$500 days to remove artifacts due to seasonal gaps and cadences aliasing. We request that the light curve time baseline covers at least three cycles to minimize false positives due to quasars' stochastic red noise variability \citep{Vaughan2016}. 

To reject spurious detections caused by noise, we fit a sinusoidal model whose frequency corresponds to the highest GLS peak at $>$99.74\% significance and calculate the signal-to-noise ratio ($\xi$) of the peak power from the amplitude of the best-fit model ($A_0$) and the scatter of the residuals ($\sigma_r$) after subtracting the sinusoidal model. The S/N ratio is defined as $\xi = A_0^2/(2\sigma_r^2)$ \citep{Horne1986}. We require $\xi > 0.5$, which means that the signal should be higher than the noise level.

\subsubsection{Other Models for Quasar Stochastic Variability}

Previous work has found deviations from the DRW model on both short ($<$1 day) and long ($>$decades) timescales \citep[e.g.,][]{Simm2016,Caplar2017,Guo2017,Smith2018}. In particular, studies based on $Kepler$ light curves suggest that a bending power law (BPL) model is needed to explain quasar light curve power spectra at the high frequency end where $f \gtrsim 1/10$ day$^{-1}$ \citep[e.g.,][]{Mushotzky2011, Edelson2013,Edelson2014}. 

To check for systematic uncertainties due to the adopted DRW model assumption, we redo the analysis using simulated light curves generated assuming the BPL model. We adopt a power spectrum index of $-3$ for f$>$1/10 day$^{-1}$ and keep the DRW model for f$<$1/10 day$^{-1}$ using \textit{pyLCSIM}\footnote{\url{http://pabell.github.io/pylcsim/html/index.html}}. We have verified that our candidate periodic quasars are robust independent of the model chosen for the simulated light curves. The significance defers by $<$0.2\% between the DRW and BPL assumptions. Our results suggest that for the periodicity window considered (i.e., 500 days--$\sim$6 years), the DRW model is appropriate to model the stochastic component of the variability.

The CAR(2, 1) model \citep{Kelly2014}, i.e., a damped harmonic oscillator, is often used to describe a periodic signal \citep{Graham2015,Moreno2019}. We have tested that the significance of the periodic signal decreases if we assume the CAR(2, 1) model for the ``stochastic'' component instead of a DRW. However, this further supports that the light curve is indeed periodic. We choose DRW to describe the stochastic component in our baseline noise models in order to separate it from any additional periodic component. The DRW model is more appropriate than CAR(2,1) in describing the stochastic variability for the general quasar population, i.e., the majority of our parent quasar sample.

\subsection{Auto-Correlation Function Analysis}\label{sec:acf}

We verify the periodic quasar candidates selected using the GLS periodograms with auto-correlation function (ACF) analysis. ACF, which calculates the correlation of a signal with a delayed copy of itself, is commonly used for identifying periodic signals in time series. We adopt the z-transformed discrete correlation function \citep[ZDCF,][]{ZDCF} method. It does not require smooth light curves and provides errors on the estimates. 

The signal from periodically driven stochastic systems is expected to be an exponentially decaying cosine function \citep{Jung1993}. We fit the ACF with the exponentially decaying cosine function 
\begin{equation}\label{eq:acf}
ACF(\tau) = Acos(\omega \tau) exp(−\lambda \tau),
\end{equation}
where $\tau$ is the time lag, $\omega$ is the frequency, and $\lambda$ is the decay rate.
We request that the best-fit ACF period is consistent with the GLS-periodogram-determined period, where the period difference is within 1$\sigma$ in at least two bands.

\begin{table}
 \caption{Summary of the numbers of candidate periodic quasars that satisfy the cumulative selection criteria .}
 \label{tab:number_candidates}
 \begin{tabular}{lc}
  \hline
  Selection Criterion & Number of Quasars \\
  \hline
  Parent Sample & 625\\
  1. $>$99.74\% in GLS periodograms & 14\\
  2. S/N ratio $\xi$ > 0.5 & 7\\ 
  3. Consistent period in ACF & 5 \\
  \hline
 \end{tabular}
\end{table}

\begin{table*}
 \caption{Basic properties of the five candidate periodic quasars. While J0252 has been presented by \citep{Liao2020}, we include it here for completeness. Listed from left to right are the source name in J2000 coordinates, spectroscopic redshift from the SDSS quasar catalog \citep{SDSS_DR14Q}, the co-added imaging PSF magnitudes in the $griz$ bands, the total numbers of SDSS and DES observations in the $griz$ bands, the estimated virial BH masses and their 1$\sigma$ statistical uncertainty (see \S \ref{sec:bhmass} for details), the estimated binary separation, the inferred gravitational wave inspiral timescale assuming $q=0.11$, and the estimated 3$\sigma$ upper limit of the radio loudness $R$ (see \S \ref{subsubsec:jet} for details) . 
 }
 \label{tab:info}
 \addtolength{\tabcolsep}{-1pt}
 \begin{tabular}{cccccccccccccc}
  \hline\hline
 Name  & Redshift & $m_{g}$ & $m_{r}$ & $m_{i}$ & $m_{z}$ & $N_{g}$ & $N_{r}$ & $N_{i}$ & $N_{z}$ & log($\frac{M_{\textrm{BH}}}{M_{\odot}}$) & $r_{\textrm{sep}}$ (pc) & $t_{\textrm{gw}}$ (yr) & $R$ \\
  \hline
  J024613.89$-$004028.2 & 1.736 & 22.60 & 22.53 & 22.37 & 22.29 & 192 & 192 & 203 & 198 & 8.74$\pm$0.07 & 0.0043 & 1.7$\times10^4$ & $<$316\\
  J024703.24$-$010032.0 & 2.525 & 20.55 & 20.33 & 20.21 & 19.93 & 201 & 212 & 216 & 221 & 8.60$\pm$0.02 & 0.0043 & 3.5$\times10^4$ & $<$77\\
  J024944.66$-$000036.8 & 1.295 & 21.40 & 21.04 & 21.01 & 21.00 & 214 & 223 & 224 & 223 & 9.00$\pm$0.04 & 0.0060 & 8.5$\times10^3$ & $<$93\\
  J025214.67$-$002813.7 & 1.530 & 21.01 & 20.60 & 20.45 & 20.47 & 212 & 223 & 222 & 227 & 8.40$\pm$0.10 & 0.0044 & 1.5$\times10^5$ & $<$39 \\
  J025406.26$+$002753.7 & 1.765 &  21.07 & 20.80 &  20.50 & 20.37 & 218 & 226 & 223 & 225 & 8.92$\pm$0.04 & 0.0058 & 1.5$\times10^4$ & $<$60 \\
  \hline
  \end{tabular}
\end{table*}

\section{Results}\label{sec:result}

\subsection{Candidate Periodically Variable Quasars}\label{subsec:result_periodicity}

Using the method described in \S \ref{sec:method}, we select five candidate periodically variable quasars from a parent sample of 625. They are: J024613.89$-$004028.2, J024703.24$-$010032.0, J024944.66$-$000036.8, J025214.67$-$002813.7, and J025406.26$+$002753.7 (hereafter J0246, J0247, J0249, J0252, and J0254 for short). We refer to them as ``candidate'' since there may be up to two false positives among the five due to noise (discussed below in \S \ref{subsec:result_significance}). \autoref{tab:number_candidates} summarizes the numbers of candidates that satisfy each selection criterion. \autoref{tab:info} lists the basic properties of the five candidate periodic quasars. Among the five candidates, \citet{Liao2020} has presented the case of J0252 as the most significant candidate and the first known case whose light curve characteristics strongly prefers hydrodynamic variability. We focus on the other four candidates for the rest of the present paper. 

\autoref{fig:lc_and_periodogram} shows the SDSS and DES light curves and the GLS periodograms of the four candidate periodic quasars. By combining SDSS and DES the total time baseline spans more than four cycles of the candidate periodicity. \autoref{tab:parameters} lists the periodicity measurements for all four candidates in the $griz$ bands, as well as the ACF periods and multi-band GLS periodogram. \autoref{fig:lc_and_periodogram} also shows archival light curves from publicly available data. While the sensitivities are lower, these archival data provide independent verification of the baseline observations from the SDSS and DES and partially fill their cadence gaps. The GLS periodograms show that the four candidate periodic quasars have observed periodicity of $\sim$3--5 years (corresponding to rest-frame $\sim$1--2 years). 

In addition to the most significant periodogram peaks, two of the four candidates (i.e., J0249 and J0254) also show significant periodogram peaks at observed $\sim$1.5 years which are just above our 500 days threshold. It is difficult to assess the robustness of these 1.5-year periodogram peaks; the other two candidates also show weak 1.5-year peaks suggesting that they may be subject to the aliasing effects caused by the cadence and seasonal gaps. We proceed focusing our discussion on the $\sim$3--5 (observed) year periodicity detection.

\begin{figure*}
\centering
\includegraphics[width=0.49\textwidth]{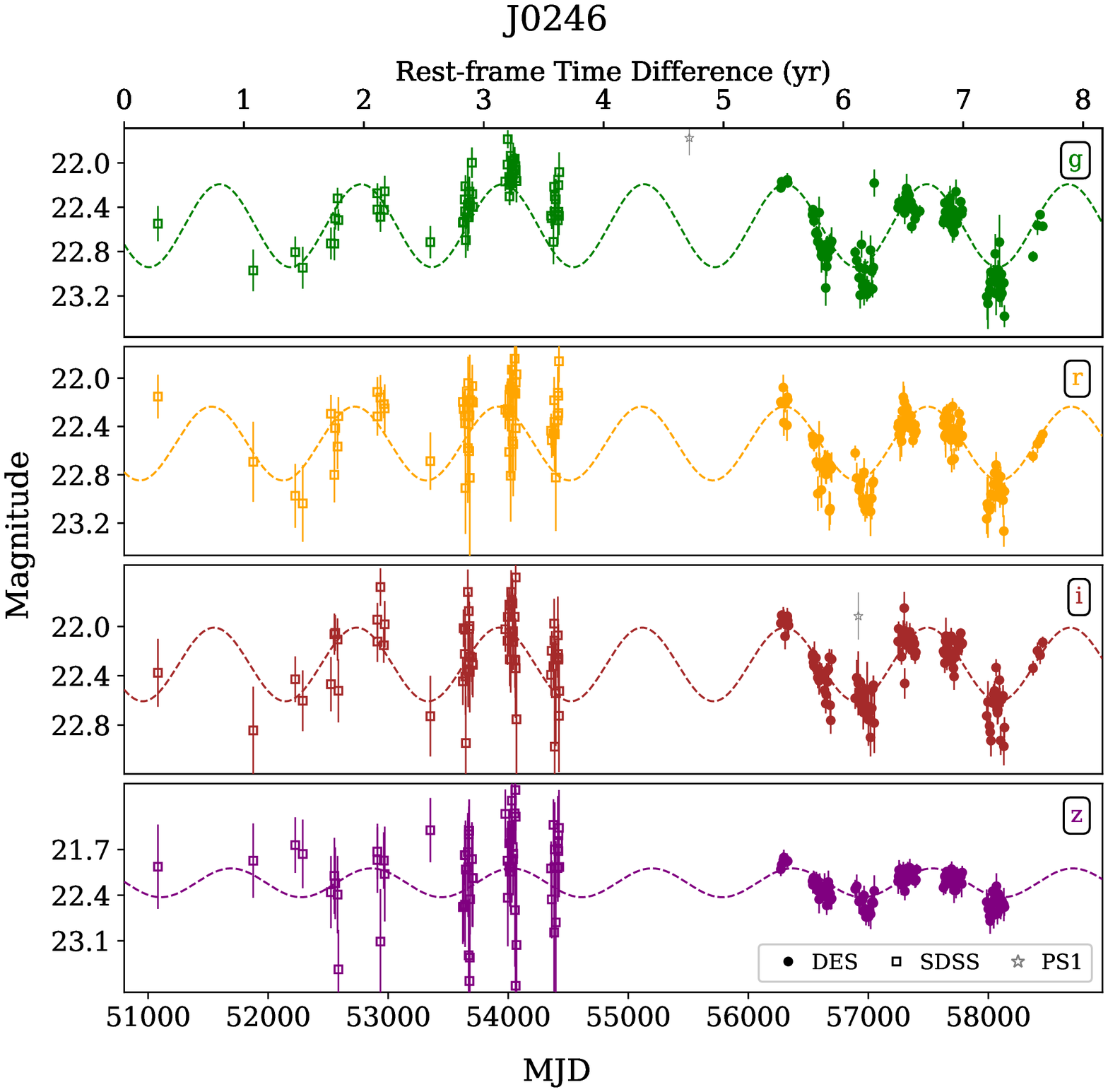}
\includegraphics[width=0.49\textwidth]{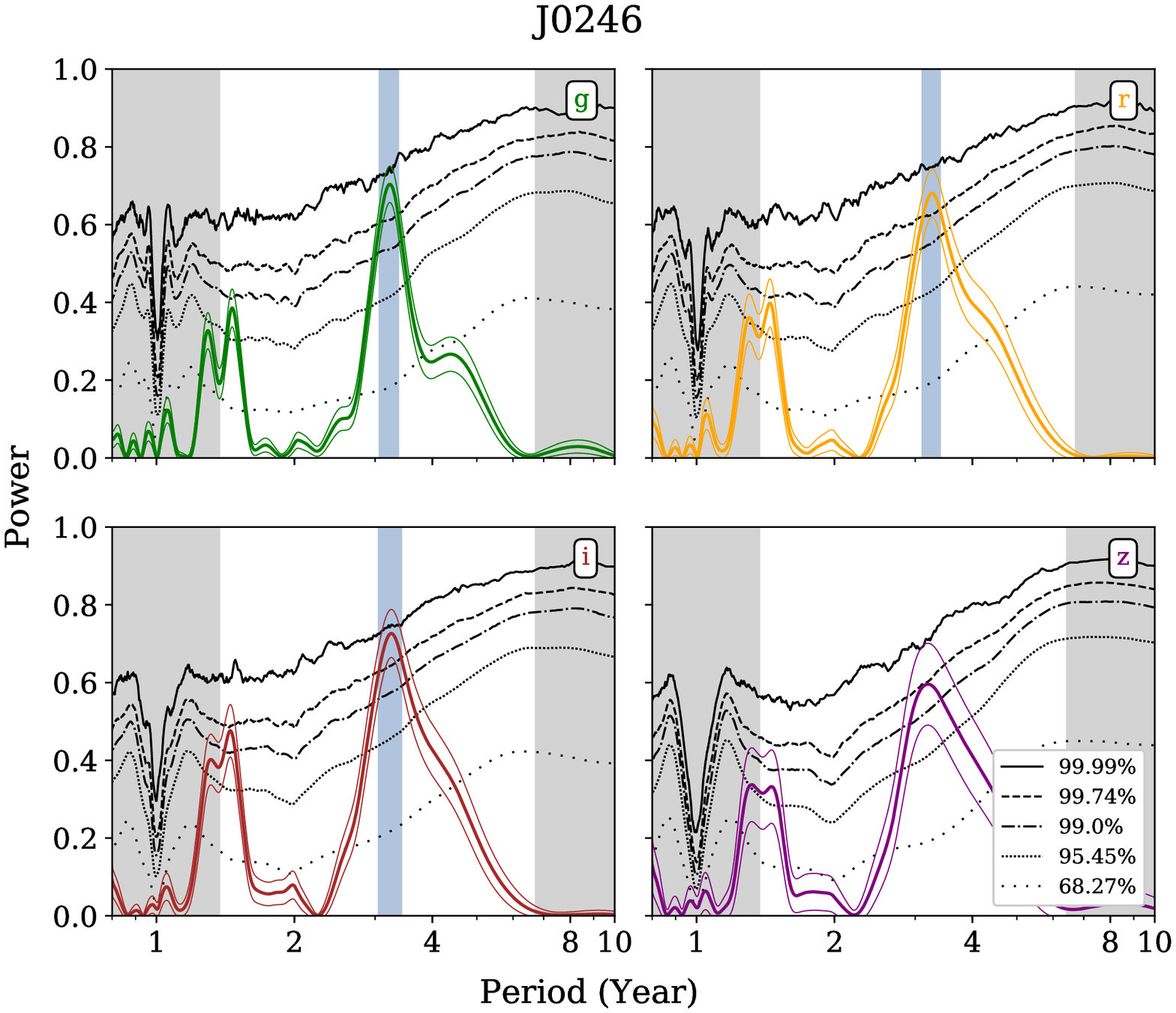}
\includegraphics[width=0.49\textwidth]{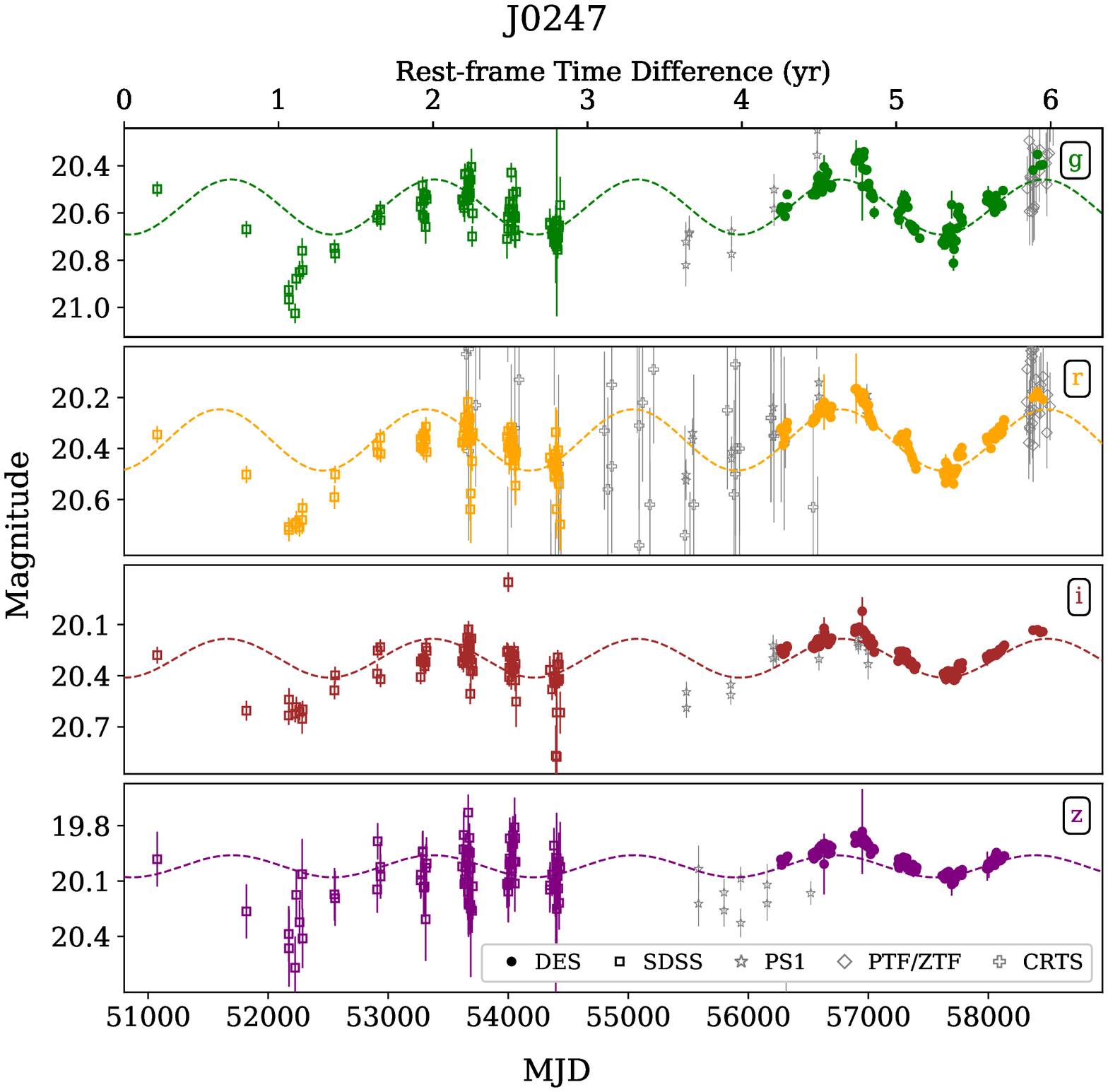}
\includegraphics[width=0.49\textwidth]{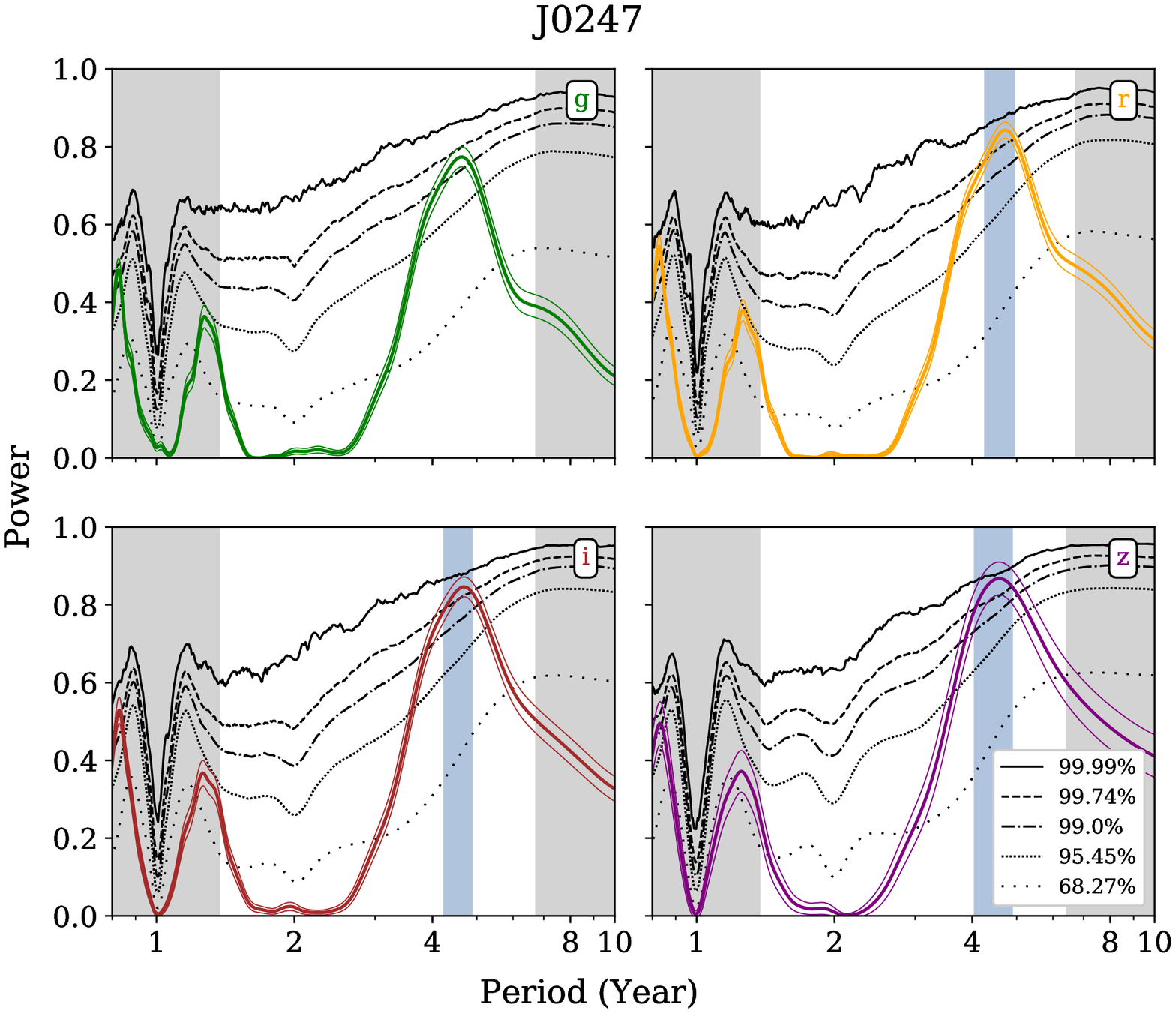}
\caption{Multi-band light curves (left column) and the GLS periodograms (right column) of the four candidate periodic quasars. \textit{Left:} New DES observations are shown as colored circles. The archival SDSS data (corrected to be on the DES system) are shown as colored squares. Gray symbols denote other publicly available archival data (CRTS data as crosses, PS1 data as stars, and PTF/ZTF data as diamonds). Error bars represent 1$\sigma$ statistical uncertainties. The dashed curves represent best-fit sinusoidal models for illustration purposes only. The residual (i.e., data $-$ model) is not supposed to be white noise because of quasars' intrinsic red noise stochastic variability. \textit{Right:} The thick curves in color show the GLS periodogram whereas the thin curves show their 1$\sigma$ errors estimated from Bootstrap re-sampling. The black curves represent the 68.27\%, 95.45\%, 99.00\%, 99.74\%, and 99.99\% significance levels calculated from 50,000 mock light curves simulated using DRW models with tailored variability parameters for each quasar. The small periods at $<$500 days and the large periods with fewer than three cycle covered by the time baseline of the observations are excluded (grey shaded regions) from our periodicity detection. The blue shaded regions indicate the periodicity uncertainty estimated using ranges above the $>$99.74\% significance.
\label{fig:lc_and_periodogram}}
\end{figure*}

\begin{figure*}
\centering
\includegraphics[width=0.49\textwidth]{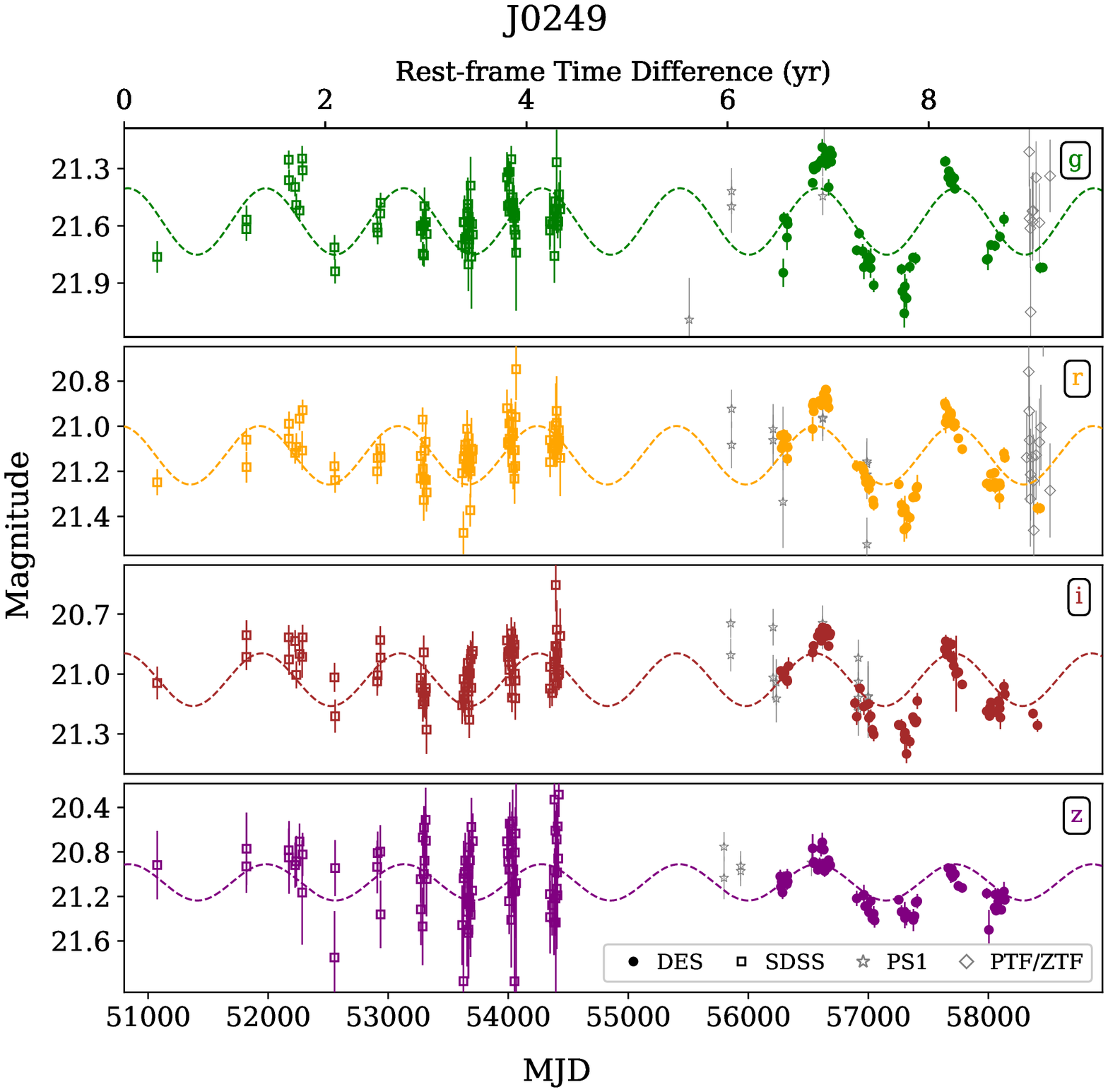}
\includegraphics[width=0.49\textwidth]{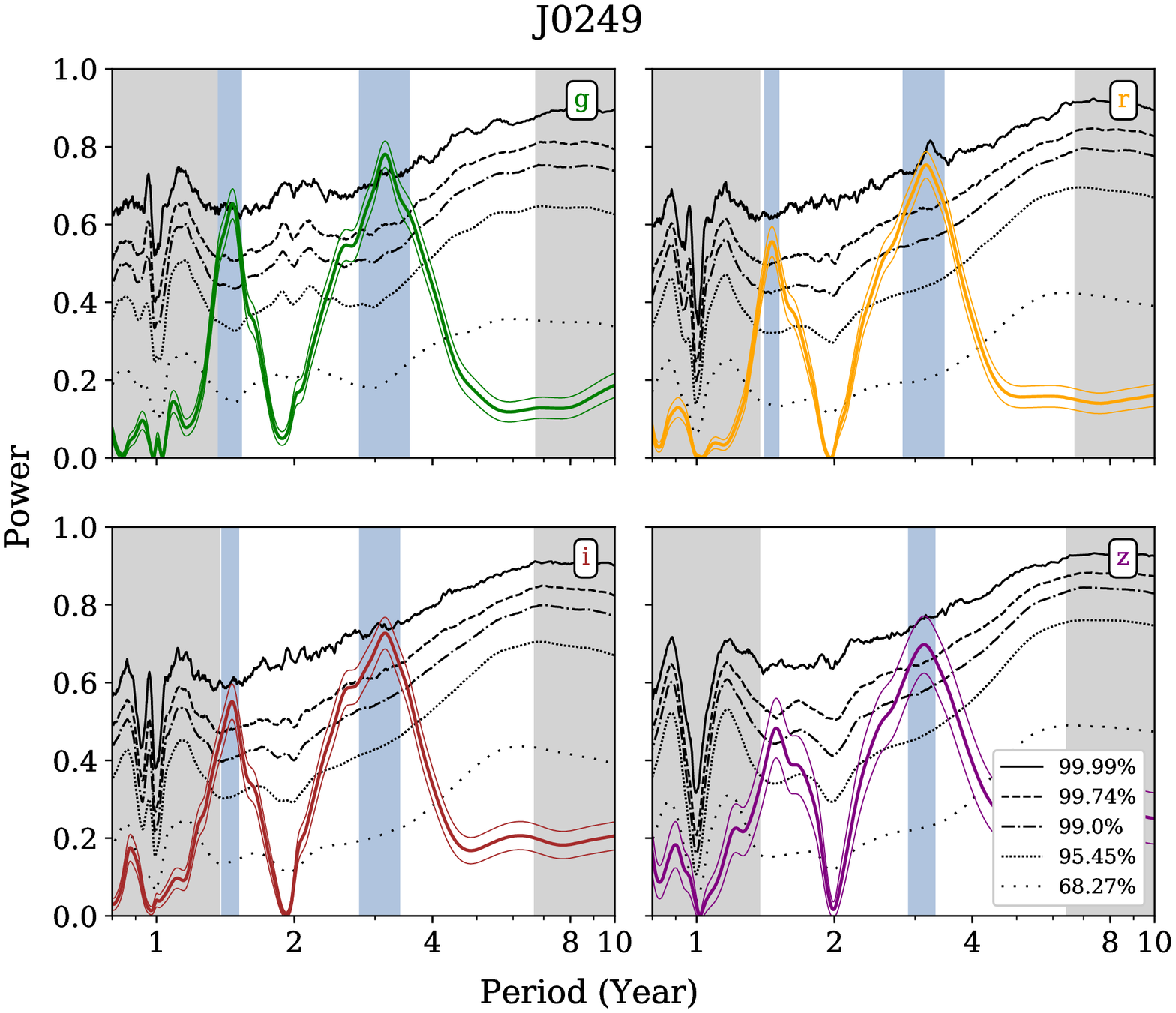}
\includegraphics[width=0.49\textwidth]{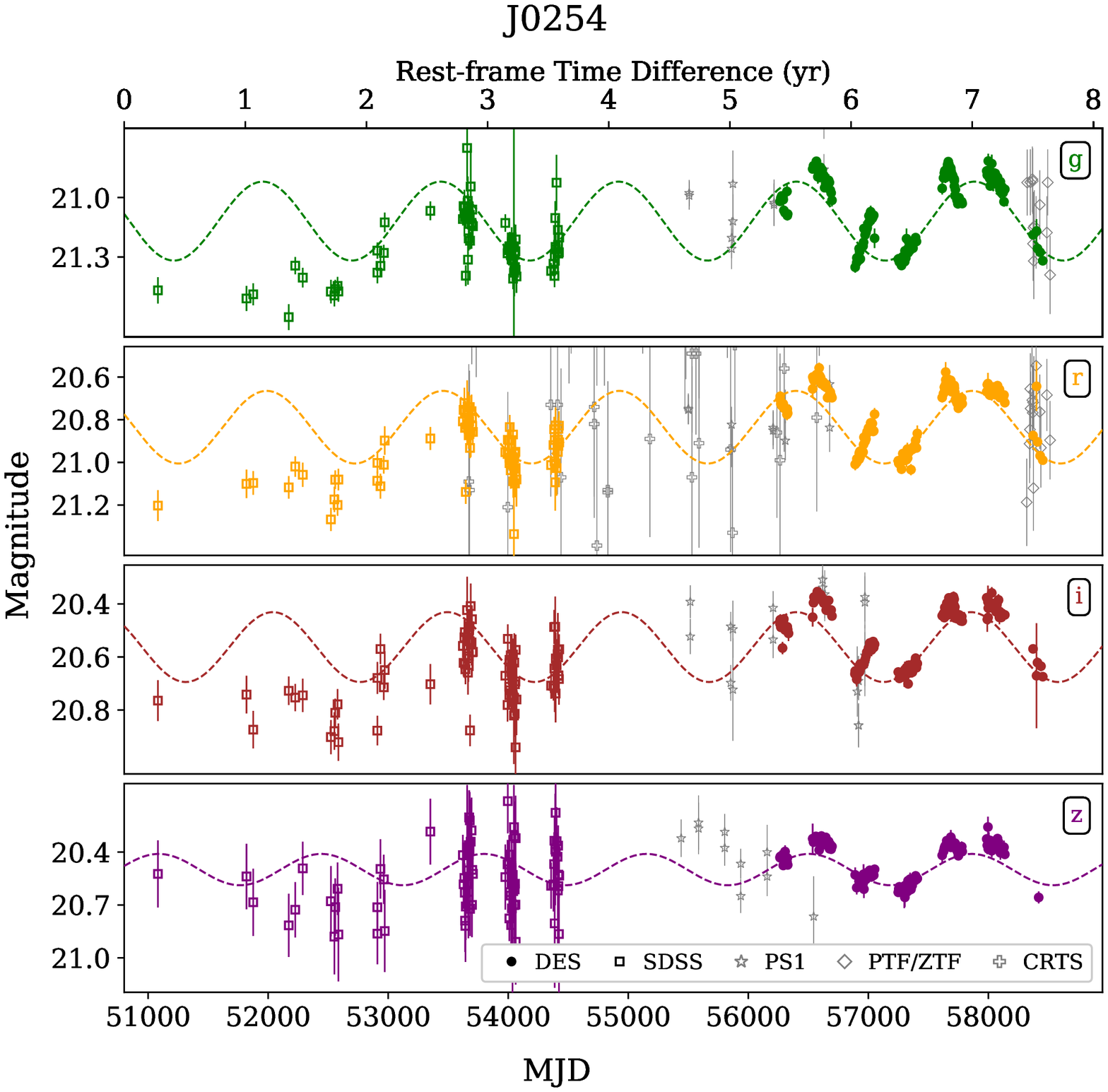}
\includegraphics[width=0.49\textwidth]{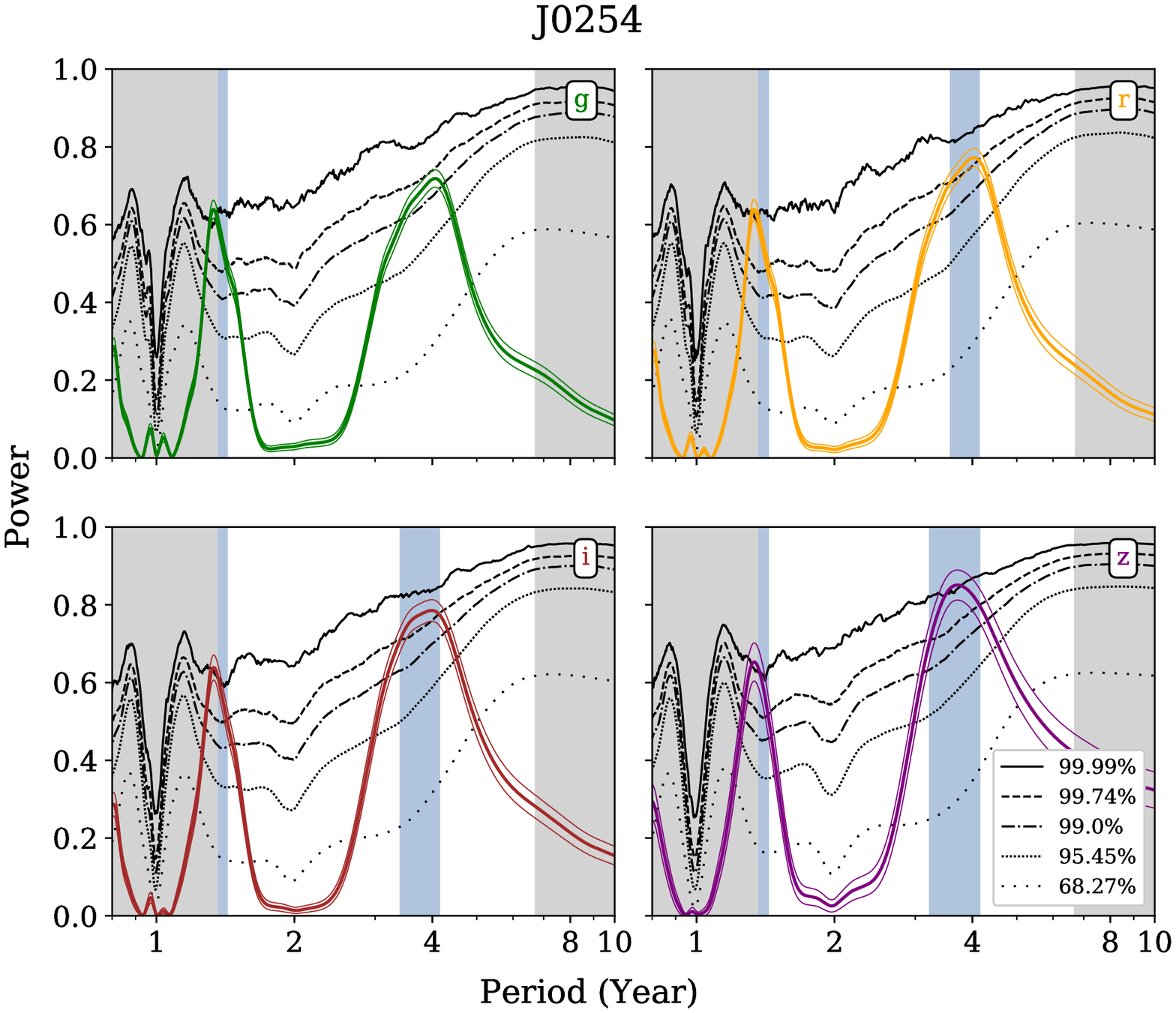}
\contcaption{}
\label{fig:tab:continued}
\end{figure*} 


\begin{table*}
 \caption{Model comparison and parameter estimation of the four candidate periodic quasars. (1): Spectral index from fitting a broken power-law model to the SDSS continuum spectrum, $\alpha_{\nu} \equiv dln(F_{\nu})/dln(\nu)$. Errors represent 1$\sigma$ uncertainties estimated from 100 MCMC simulations. (2): Variability amplitude from the best-fit sinusoidal model. (3): The total time span of the observations. (4): Period from the highest peak in the GLS periodogram. Errors represent 1$\sigma$ uncertainties estimated by 1000 perturbed light curves. (5): Period found in the ACF fitting. Errors represent 1$\sigma$ uncertainties. (6): Period from the highest peak in the multi-band Lomb-Scargle periodogram. Errors represent 1$\sigma$ uncertainties estimated by 1000 perturbed light curves. (7): The single-frequency p-value empirically estimated by the fraction of simulated light curves that pass all three criteria at the detected peak frequency of each candidate. (8): The global p-value empirically estimated by the fraction of simulated light curves whose periodicity is more significant than the candidate at any given frequency within the search range \citep{Barth2018} according to all three periodicity selection criteria. (9): The best-fit correlation timescale calculated in the DRW model. (10): The best-fit variance calculated in the DRW model and $\sigma^2_{{\rm DRW}} = \tau \sigma^2/2$. (11) \& (12): Bayesian information criterion (BIC) differences between a periodic model (including a sinusoidal model expected from Doppler boosting model, and one bursty accretion models assuming mass ratio of 0.11) and the null hypothesis (stochastic quasar variability characterized by a DRW model). A negative $\Delta$BIC value indicates that the periodic model is more preferred than the DRW model. $\Delta$BIC $<$ −10 suggests strong evidence.  
 }
 \label{tab:parameters}
 {\renewcommand{\arraystretch}{1.2}
 \addtolength{\tabcolsep}{-4pt}
 \begin{tabular}{cccccccccccccc}
  \hline\hline
 Name  & band  & $\alpha_{\nu}$ & $A$ & $\Delta$T & $P_{\rm GLS}$ & $P_{{\rm ACF}}$ & $P_{{\rm multi-band}}$ & $p_{{\rm single}}$ & $p_{{\rm global}}$ & $\tau_{\rm DRW}$ & $\sigma^2_{{\rm DRW}}$ & $\Delta$BIC$_{{\rm sin}-{\rm DRW}}$ & $\Delta$BIC$_{{\rm Acc}-{\rm DRW}}$  \\
 & & & (mag) & (days) & (days) & (days) & (days) & ($\times10^{-4}$) & ($\times10^{-4}$) & ($\times10^2$ days) & (mag$^2$) &  & \\
 & & (1) & (2) & (3) & (4) & (5) & (6) & (7) & (8) & (9) & (10) & (11) & (12) \\
  \hline
  \multirow{4}{3em}{J0246} & g & $-$0.8$\pm$1.1 & 0.38$\pm$0.02 & 7371 & 1178$\pm$3 & 1173$\pm$7 & \multirow{4}{3em}{1184$\pm$2} & 1 & 19 & 3.4$^{+6.3}_{-0.4}$ &  0.08$^{+0.12}_{-0.01}$ & $-$4.91 & $-$5.67  \\
  & r & $-$1.3$\pm$0.9 & 0.29$\pm$0.02 & 7371 & 1194$\pm$5 &  1187$\pm$16 & & 5 & 83 & 3.8$^{+6.6}_{-0.5}$ & 0.07$^{+0.10}_{-0.01}$ & $-$3.17 & $-$1.38 \\
  & i & $-$0.8$\pm$1.1 & 0.30$\pm$0.02 & 7371 &  1189$\pm$5 & 1177$\pm$8 & & 2 & 39 & 2.9$^{+5.3}_{-0.4}$ & 0.05$^{+0.06}_{-0.01}$ & $-$12.42 & $-$11.68 \\
  & z & $-$0.6$\pm$1.3 & 0.21$\pm$ 0.04 & 7053 & 1169$\pm$17 & 1158$\pm$46 & & 114 & 811 & 7.1$^{+10.1}_{-1.4}$ & 0.08$^{+0.10}_{-0.01}$ & $-$8.05 & $-$10.75 \\
  \hline
  \multirow{4}{3em}{J0247}& g & $-$1.61$\pm$0.10 & 0.119$\pm$0.009 & 7378 & 1693$\pm$7 & 1743$\pm$23 & \multirow{4}{3em}{1705$\pm$3} & 7 & 85 & 4.6$^{+8.3}_{-0.4}$ & 0.02$^{+0.02}_{-0.01}$ & 3.04 & 5.94 \\
  & r & $-$0.53$\pm$0.09 & 0.121$\pm$0.008 & 7378 & 1723$\pm$6 & 1748$\pm$23 & & 1 & 28 & 11.1$^{+9.3}_{-2.1}$ & 0.02$^{+0.01}_{-0.01}$ & 1.06 & 2.15\\
  & i & $-$0.75$\pm$0.16 & 0.123$\pm$0.011 & 7378 & 1709$\pm$8 & 1733$\pm$35 & & 4 & 43 & 15.9$^{+5.9}_{-4.8}$ & 0.015$^{+0.005}_{-0.005}$ & 2.65 & 0.06 \\
  & z & $-$0.97$\pm$0.26 & 0.061$\pm$0.012 & 7059 & 1672$\pm$17 & 1485$\pm$79 & & 21 & 218 &  19.4$^{+1.7}_{-8.7}$ & 0.007$^{+0.002}_{-0.003}$ & 1.91 & 2.32 \\
  \hline
  \multirow{4}{3em}{J0249} & g & $-$1.91$\pm$0.29 & 0.178$\pm$0.019 & 7378 & 1150$\pm$3 &  1133$\pm$59 & \multirow{4}{3em}{1153$\pm$2} & 1 & 7 & 2.3$^{+3.1}_{-0.3}$ & 0.03$^{+0.03}_{-0.01}$ & 2.18 & 3.49   \\
  & r & $-$1.49$\pm$0.10 & 0.129$\pm$0.013 & 7356  & 1157$\pm$4 & 1205$\pm$29 & & 1 & 21 & 2.9$^{+4.5}_{-0.3}$  & 0.01$^{+0.03}_{-0.01}$ & 4.50 & 2.01\\
  & i & 0.86$\pm$0.24 & 0.134$\pm$0.014 & 7334 & 1152$\pm$4 & 1149$\pm$46 & & 1 & 30 & 3.2$^{+5.1}_{-0.4}$ & 0.02$^{+0.02}_{-0.01}$ & 2.55 & $-$0.08\\
  & z & $-$0.64$\pm$0.85 & 0.164$\pm$0.030 & 7061 & 1147$\pm$9 & 1116$\pm$54 & & 30 & 268 & 4.5$^{+9.0}_{-1.4}$ & 0.03$^{+0.04}_{-0.01}$ & 1.11  & $-$1.25\\
  \hline
  \multirow{4}{3em}{J0254} & g & $-$1.52$\pm$0.17 & 0.210$\pm$0.013 & 7372 & 1482$\pm$4 & 1476$\pm$29 & \multirow{4}{3em}{1467$\pm$3}  & 2 & 67 & 9.4$^{+9.9}_{-1.5}$ & 0.04$^{+0.04}_{-0.01}$ & 4.70 & 0.14 \\
  & r & $-$0.028$\pm$0.10 & 0.180$\pm$0.012 & 7372 & 1469$\pm$5 & 1498$\pm$19 &  & 2 & 32 & 12.5$^{+8.4}_{-2.9}$ & 0.03$^{+0.01}_{-0.01}$ & 0.36 & 1.72 \\
  & i & $-$1.29$\pm$0.12 & 0.134$\pm$0.010 & 7372 & 1454$\pm$11 & 1476$\pm$19 & & 3 & 48 &  11.0$^{+9.7}_{-2.1}$ & 0.02$^{+0.01}_{-0.01}$ & 1.74 & 2.99 \\
  & z & 0.03$\pm$0.22 & 0.090$\pm$0.014 & 7338 & 1353$\pm$13 & 1431$\pm$40 & & 3  & 40 & 8.0$^{+10.1}_{-1.4}$ & 0.01$^{+0.01}_{-0.01}$ & $-$2.54 & $-$3.63\\
  \hline
  \end{tabular}}
\end{table*}





\subsection{Statistical Significance of the Candidate Periodic Quasars as a Population}\label{subsec:result_significance}

We find five candidate periodic quasars in a sample of 625 spectroscopically confirmed quasars in a 4.6 deg$^2$ field. To understand the statistical significance of these periodic candidates as a population, we calculate the global false alarm probability (FAP) taking into account the ``look-elsewhere effect'' \citep[e.g.,][]{Gross2010}. 
We estimate the approximate FAP using the effective number of independent frequencies N$_{{\rm eff}}$, where N$_{{\rm eff}}$ is calculated by dividing the observed frequency window $\Delta f_{{\rm obs}}$ (ranging from $\sim$1/7 year$^{-1}$ to 1/500 day$^{-1}$ for the five candidates) by the expected peak width, $\delta f$. We estimate $\delta f$ from the peaks of the candidates, which are $\sim$0.20 year$^{-1}$. The FAP is estimated as \citep[e.g.,][]{VanderPlas2018}
\begin{equation}
    {\rm FAP} \sim 1 - [ P_{{\rm single}}]^{N_{{\rm eff}}}.
\end{equation}
Under the GLS periodogram selection criterion only, $P_{{\rm single}}$=99.74\% and N$_{{\rm eff}}\sim$3, we therefore expect to see $\sim$five false positives from a sample of 625 with an ${\rm FAP}\sim 0.8\%$. However, we find 14 candidates that satisfy the GLS periodogram selection criterion (\autoref{tab:number_candidates}), which are significantly more than the expected five. This suggests that we are not just seeing noise in the detected candidates. Since we have also adopted two other selection criteria (the ACF and the S/N), the expected number of false positives should be fewer than five and the expected FAP by combining all three criteria should be $<5/14$. We therefore estimate that up to $\sim5/14$ of the five periodic candidates (i.e., $\sim$two objects) may be false positives caused by red noise. 

We also empirically estimate the global FAP using the 50,000 simulated light curves for each quasar following \citet{Barth2018}. This global FAP combines all three tests and takes into account the look-elsewhere effect \citep{Gross2010} by counting for all possible false positives within the searched frequency range. \autoref{tab:number_candidates} lists the empirically estimated global p-value for each candidate. The expected number of false positives estimated using the global FAP is ${\sim}1.2{\pm}0.6$ (adopting the band with the most significant detection for each candidate), which is similar to our other estimate as discussed above.

\subsection{Black Hole Mass Estimation}\label{sec:bhmass}

All four candidate periodic quasars have optical spectra available in the SDSS DR16 \citep{SDSS_DR16} data archive. They were all observed using the BOSS spectrograph by the SDSS-III/BOSS \citep{Dawson2013} or SDSS-IV/eBOSS survey \citep{Dawson2016}. The BOSS spectra cover observed 3650--10400 {\rm \AA} with a spectral resolution of $R=$1850--2200. 

To measure the broad emission lines for virial black hole mass estimate, we fit the SDSS spectra following the procedures described in \cite{Shen2012} using the code PyQSOFit \citep{PyQSOFit,Shen2019}. The spectral model consists of a power-law continuum, a pseudo-continuum constructed from the Fe II emission templates, and single or multiple Gaussian components for the narrow and broad emission lines. We also fit a broken power-law model to the continuum to obtain the separate spectral index $\alpha_{nu}$ at each band for the Doppler boosting hypothesis test (see \S \ref{sec:doppler} below). \autoref{fig:spectra} shows the SDSS spectra and our best-fit models of the four candidate periodic quasars.

Since the size of the broad line region (BLR) is likely to be much larger than the expected binary separation, the BLR gas would see a binary black hole as a single source. We estimate the total BH mass using the single-epoch estimator assuming the BLR gas clouds are virialized \citep{Shen13_BHmass}. The virial BH mass is estimated by
\begin{equation}
    \text{log}_{10} \bigg(\frac{M_{\text{BH}}}{M_{\odot}}\bigg) = a + b \text{ log}_{10}\bigg(\frac{\lambda L_{\lambda}}{10^{44}\,\text{erg s}^{-1}}\bigg) + 2 \text{ log}_{10}\bigg(\frac{\text{FWHM}}{\text{km s}^{-1}}\bigg),
\end{equation}
where $L_\lambda$ is the monochromatic continuum luminosity at the wavelength $\lambda$, FWHM is the full width at half maximum of the broad emission line, and the coefficients $a$ and $b$ are empirically calibrated against local AGNs with Reverberation Mapping (RM) masses and internally with other lines. We adopt \MgII\ as the primary BH mass estimator and \CIV\ as the secondary estimator if \MgII\ is not covered or too noisy. \MgII\ is generally considered to be more reliable than \CIV\ for BH mass estimation \citep{Shen13_BHmass}, given that \CIV\ is likely to be more affected by outflows and the larger scatter between \CIV\ and \hbeta\ masses observed in high-redshift quasars \citep{Shen2012}. We adopt the calibration coefficients $a$ and $b$ by \citet{Vestergaard2006} and \citet{Vestergaard2009}. \autoref{tab:info} lists the virial BH mass estimates and their 1$\sigma$ statistical errors for the four candidate periodic quasars.

\section{Light Curve Modeling}\label{sec:modeling}

\subsection{Light Curve Model Comparison and Parameter Estimation}\label{subsec:lc_model}

We have selected five candidate periodic quasars in which the periodicity is unlikely to be caused by stochastic red noise based on the GLS periodograms calibrated using tailored simulations. The periodic signals found in the GLS periodograms and ACF analysis may have various origins. We focus our discussion on the binary scenario first. 

The shape of the light curves may offer important clues to the physical origin of any periodicity. A sinusoidal shape is expected from Doppler boosting \citep[e.g.,][]{DOrazio2015,Charisi2018}, whereas a more bursty, sawtooth pattern is expected from hydrodynamic variability of circumbinary accretion  \citep{Farris2014,Duffell2020}. 

As an alternative test, here we compare different light curve models to assess if an additional periodic signal is needed to explain the light curve on top of a stochastic background. We adopt a maximum likelihood approach for the model comparison and parameter estimation. We use the Bayesian information criterion (BIC), which is defined as
\begin{equation}
{\rm BIC} = -2 \ln ({\cal L}) + k \ln (N),
\end{equation}
where $\cal{L}$ is the likelihood function, $k$ is the number of free model parameters, and $N$ is the number of data points. The likelihood function is given by 
\begin{equation}
\label{eq:likelihood_func}
{\cal L} \propto
\det | C |^{- \frac{1}{2}}
\exp \bigg[ - \frac{1}{2} (X_i - M_i) \left( C^{-1} \right)_{ij} (X_j - M_j) \bigg] \, ,
\end{equation}
where $C$ is the co-variance matrix, $X_i$ is the observed flux, and $M_i$ is the model flux at the observation time $t_i$. The co-variance matrix includes a correlated red noise for the stochastic red noise variability. It is given by
\begin{equation}\label{eq:covariance}
    C_{i j}=
    \begin{cases}
      {\sigma_i}^2+{\sigma}^2, & \text{if}\ i=j \\
      \sigma^2 \exp{\bigg[\frac{- |t_i - t_j |}{\tau}\bigg]}, & \text{otherwise }
    \end{cases} \, ,
\end{equation}
where $\sigma_i$ is the 1$\sigma$ measurement error at the observation time $t_i$, $\tau$ is the correlation time, and $\sigma^2$ is the variance. The off-diagonal terms correspond to a correlated red noise. The null hypothesis is a correlated red noise with a flat amplitude, the correlation time $\tau$, and the variance $\sigma^2$, which is equivalent to a DRW model. 

Similar to \citet{Liao2020}, we consider two physically different models for the periodic signal. One is a smooth sinusoidal model and the other is a more bursty sawtooth-pattern model predicted by circumbinary accretion disk simulations with a binary mass ratios of $q=0.11$ from \citet{Farris2014}. The light curve models of \citet{Farris2014} were generated from two-dimensional (2D) hydrodynamical simulations of circumbinary disk accretion using the finite-volume code {\it DISCO} \citep{Duffell2016}. {\it DISCO} solves the 2D viscous Navier-Stokes equations on a high-resolution moving mesh, which reduces the advection error compared to that from a fixed mesh. The \citet{Farris2014} model was the first 2D study including the inner cavity, using shock-capturing thin disks over the viscous timescales. The simulations last longer than a viscous time so that the solutions represent a quasi-steady accretion state. We adopt the mass ratio $q=0.11$ because it represents a characteristic regime in the light-curve behaviors (see Figure 9 of \citealt{Farris2014}) where a strong peak is seen in the periodograms corresponding to the orbital frequency of an over-dense lump in the accretion disk. We do not attempt to further discriminate between models of different mass ratios in view of significant uncertainties in both the models and the data. We assume constant mass-to-light ratio, though we acknowledge that the relation between observed luminosity and the accretion rate is not a simple mapping and is subject to more complicated factors such as radiative transfer.



The sinusoidal model has six free parameters: the red noise amplitude, the red noise correlation time, period, phase, amplitude, and average flux. The bursty model also has six free parameters: the red noise amplitude, the red noise correlation time, period, phase shift, amplitude of variation, and the flux zero point. The pure stochastic red noise model (i.e., the null hypothesis) has three free parameters: the red noise amplitude, the red noise correlation time, and the mean flux. A lower BIC value indicates that the model is preferred. The proportional constant in \autoref{eq:likelihood_func} is set to unity. 

Based on calculating the BIC differences, \citet{Liao2020} have shown for the case of J0252 that the periodic models are strongly preferred over a purely stochastic red noise model; a more bursty model is also preferred over a smooth sinusoidal model expected from Doppler boosting. This points to circumbinary accretion disk variability as the physical origin for driving the periodicity seen in J0252.

\autoref{tab:parameters} lists the BIC differences between the two periodic models and the null hypothesis (a pure DRW model) for the four candidate periodic quasars presented in this paper. Among the four candidates, J0246 has consistently negative BIC differences for both periodic models compared to the null hypothesis in all bands. In particular, the BIC differences in the $iz$ bands are $<-10$, suggesting strong evidence that the periodic models are preferred over a pure stochastic model. The BIC differences between the two periodic models are small, showing no preference for either periodic model.

For the other three candidates (i.e., J0247, J0249, and J0254), the BIC differences are small, which shows that the two periodic models are not preferred over the pure stochastic model. While this may seem at odds with our conclusion based on the GLS periodograms, it may be a result of the limitation of the specific periodic models we have considered. The two given periodic models may not best describe the periodic features seen in our candidates, besides, more free parameters in the periodic models will also have higher penalty in the BIC calculation. The best-fit BIC thus might not always favor the two periodic models for our candidates. We cannot rule out the possibility that the data may prefer other forms of periodic models over a pure stochastic model or even other forms of stochastic model.


\begin{figure*}
\centering
\includegraphics[width=0.48\textwidth]{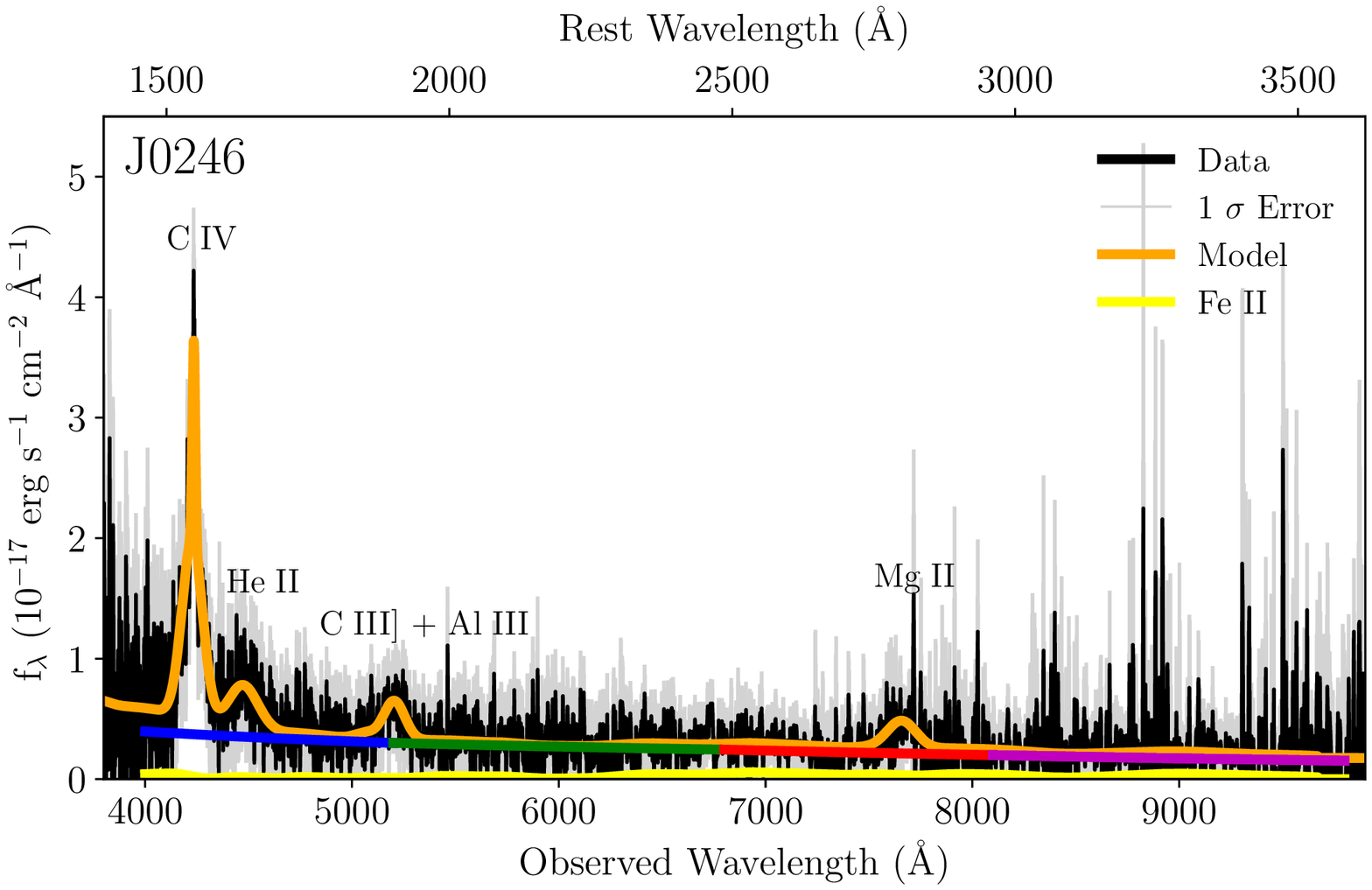}
\includegraphics[width=0.48\textwidth]{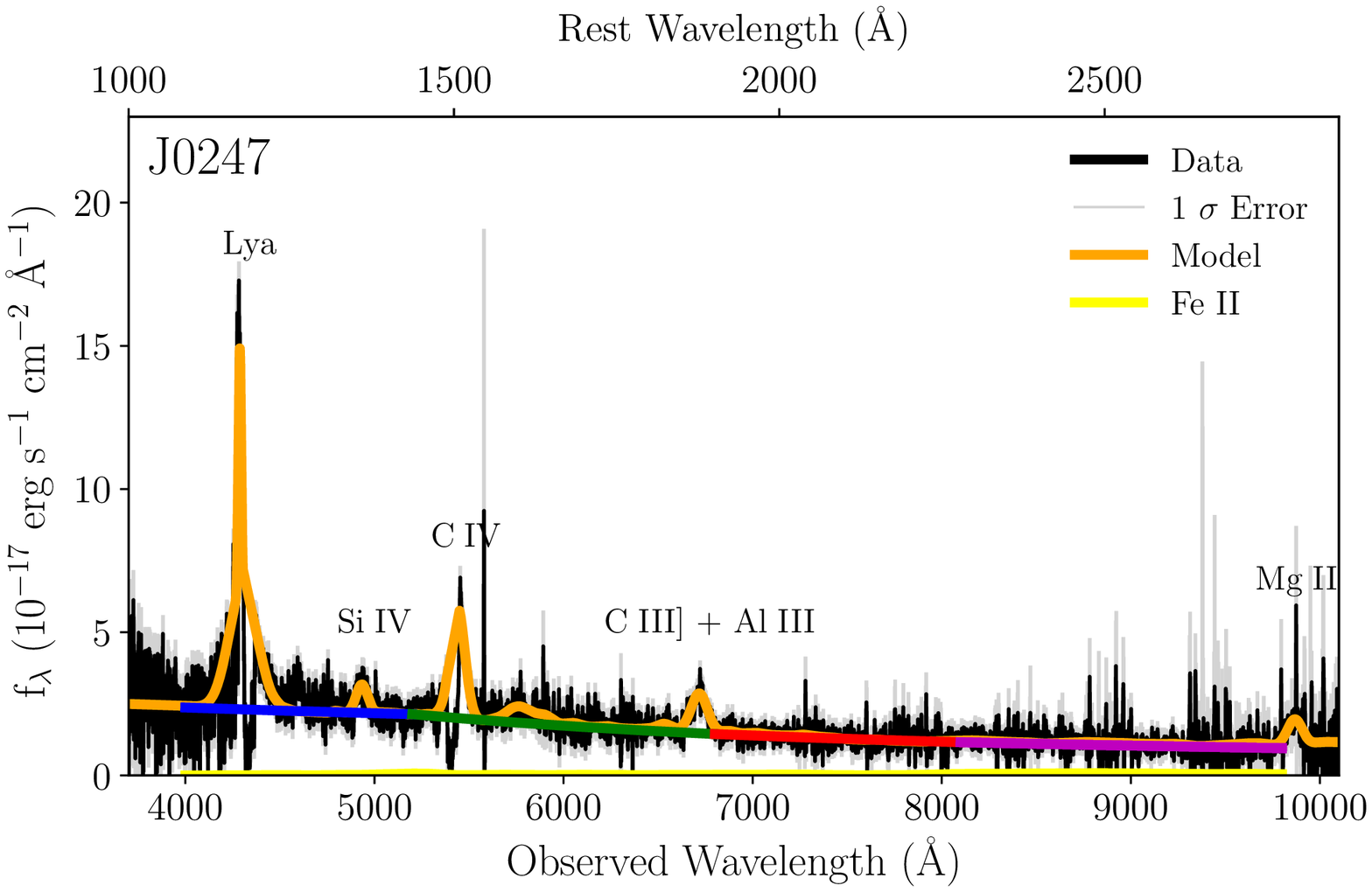}
\includegraphics[width=0.48\textwidth]{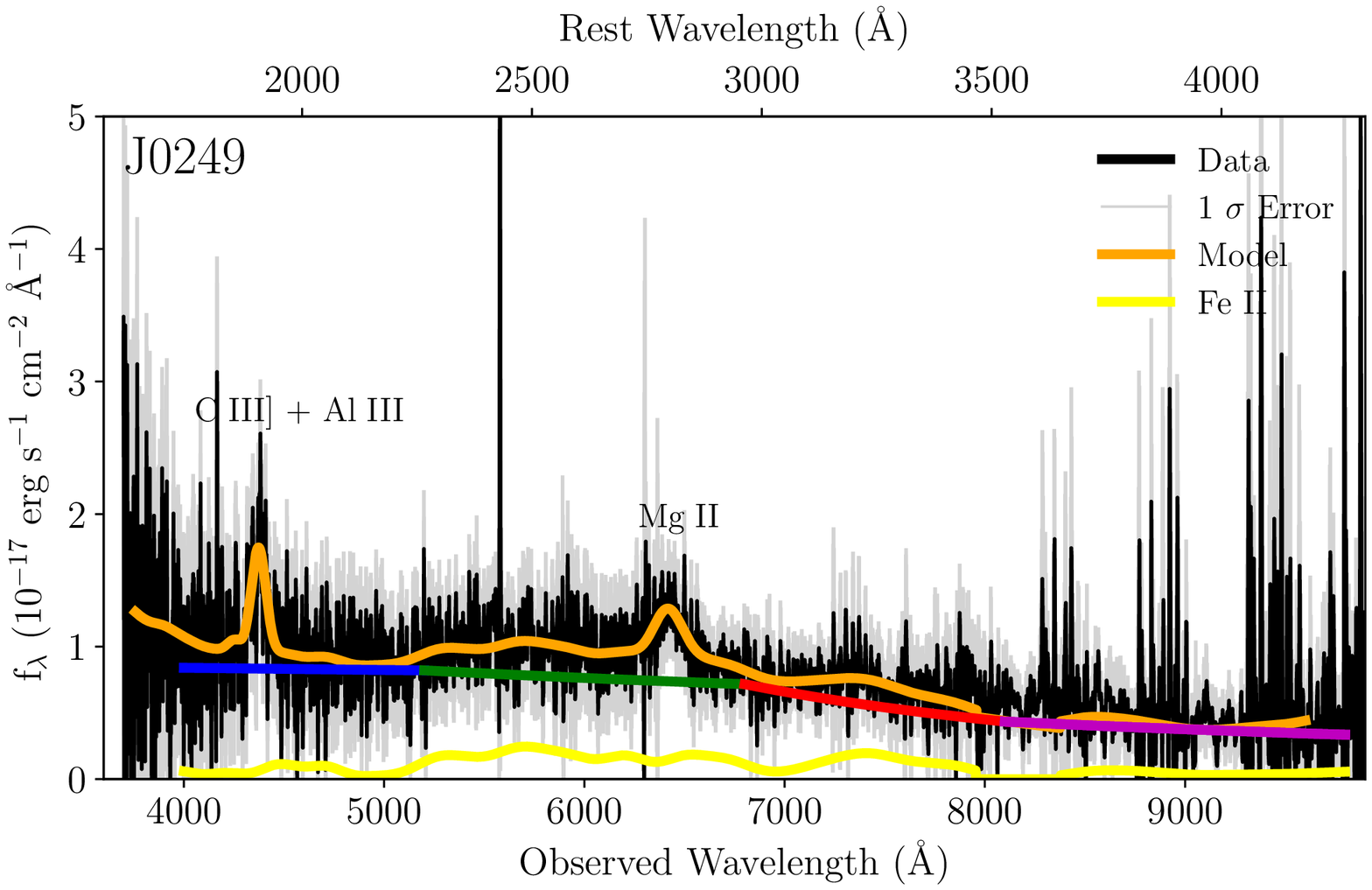}
\includegraphics[width=0.48\textwidth]{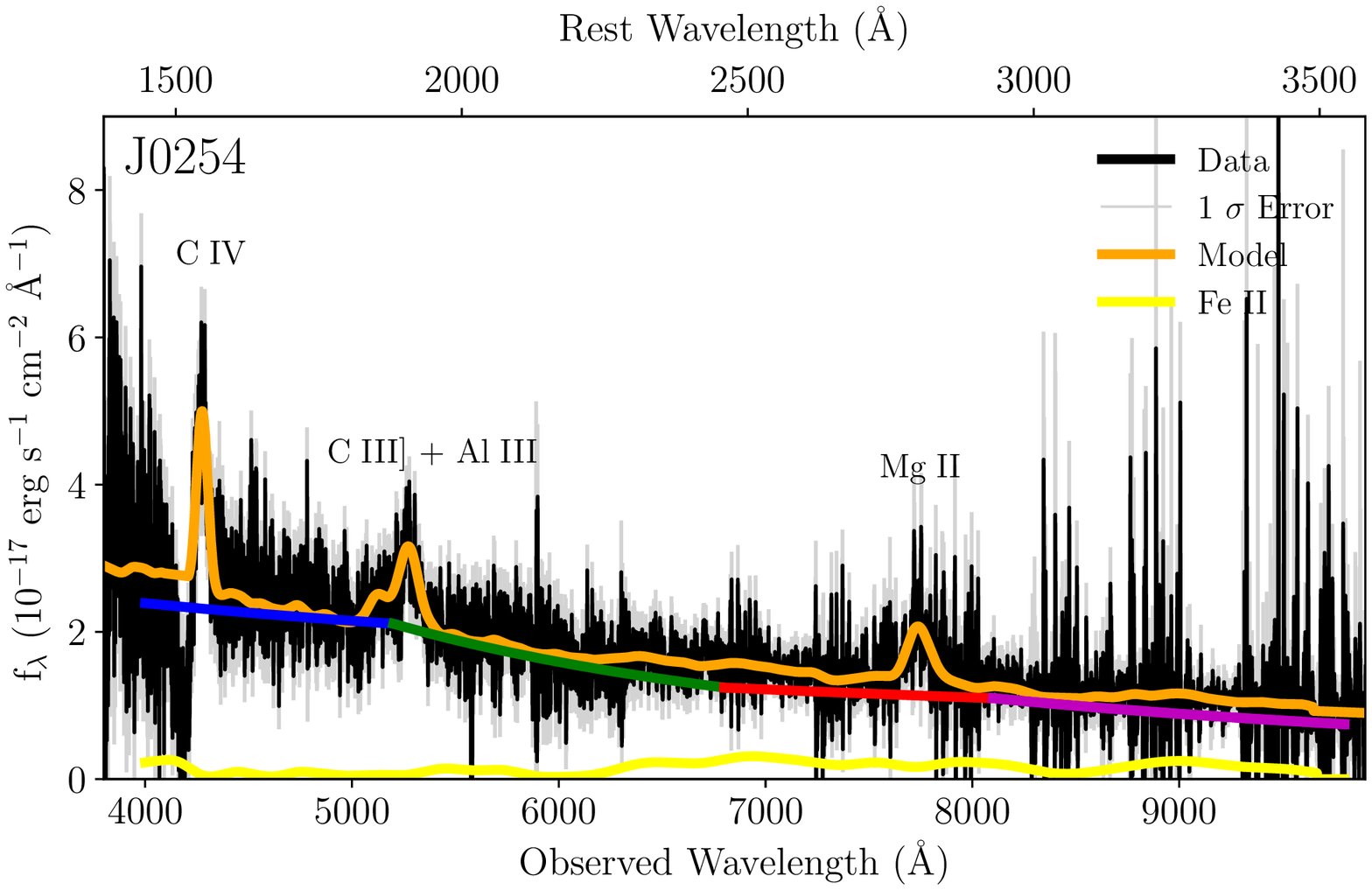}
\caption{SDSS/BOSS optical spectra of the four candidate periodic quasars. Shown are the data (black), the 1$\sigma$ rms error (gray), the best-fit model (orange), the Fe {\tiny II} pseudo-continuum (yellow), and the broken power-law model for the emission-line- and Fe {\tiny II}-subtracted continuum (with the $griz$ bands plotted in blue, green, red, and magenta, respectively).
}
\label{fig:spectra}
\end{figure*}

\subsection{Testing the Doppler Boost Hypothesis}\label{sec:doppler}

For a binary black hole in a circular orbit, the emission from the secondary black hole will vary due to Doppler boosting. We can test if the periodicity may be caused by Doppler boosting by quantifying the frequency-dependent variability amplitudes \citep{DOrazio2015,Charisi2018}. 
For an object moving relativistically, assuming the emitted radiation has a power-law spectrum $F_{\nu} \propto \nu^{\alpha_{\nu}}$, where $F_{\nu}$ is the spectral flux density at frequency $\nu$ and $\alpha_{\nu}$ is the spectral index, the observed flux is related to the emitted flux as
\begin{equation}
    F^{obs}_{\nu} = D^{3-\alpha_{\nu}} F^{em}_{\nu},
\end{equation}
where 
\begin{equation}
    D = \frac{1}{\gamma \left(1 - \textit{v}_\parallel / c\right)}, \text{ }\gamma = \sqrt{1-(\frac{\textit{v}}{c})^2},
\end{equation}
$c$ is the speed of light, \textit{v} is the orbital velocity, and $v_\parallel$ is the line-of-sight velocity. 

For a binary in a circular orbit, to the first-order approximation, the variability due to relativistic Doppler boost is
\begin{equation}
\frac{\Delta {F}_\nu}{{F}_\nu} = \left(3-\alpha_\nu\right) \frac{\textit{v}}{c} \cos\phi \sin i \ ,
\end{equation}
where \textit{v} is the orbital velocity of the more luminous black hole (typically the less massive black hole, with the primary black hole contributes negligible flux), $i$ is the inclination angle of orbital plane to the line-of-sight ($i=0$ is face-on and $i=\pi/2$ is edge-on), and $\phi$ is the phase angle. We take the orbital separation to be constant during the observations, since the observing time is much smaller than the coalescence timescale of the binary.

The relative amplitude of the periodic signal between different bands is given by
\begin{equation}
    \frac{A_i}{A_j} = \frac{3-\alpha_i}{3-\alpha_j},
\end{equation}
where $A$ is the amplitude of the periodic signal, and $i$ and $j$ denote two optical bands. By comparing the observed amplitude ratios from multi-band light curves with the theoretical values, we can test the Doppler boost hypothesis.

\begin{figure}
\centering
\includegraphics[width=0.23\textwidth]{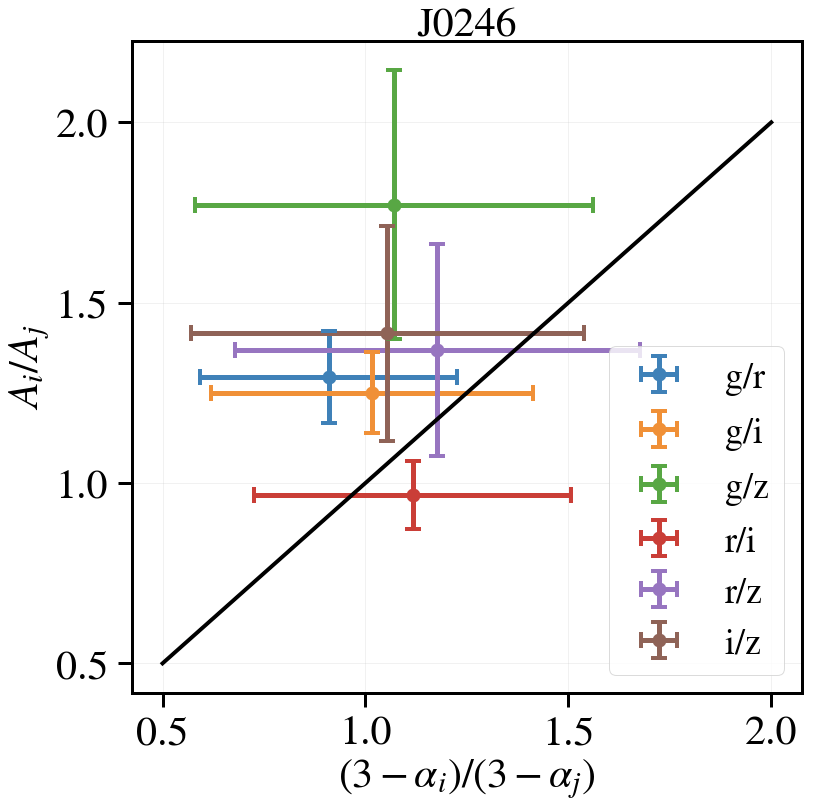}
\includegraphics[width=0.23\textwidth]{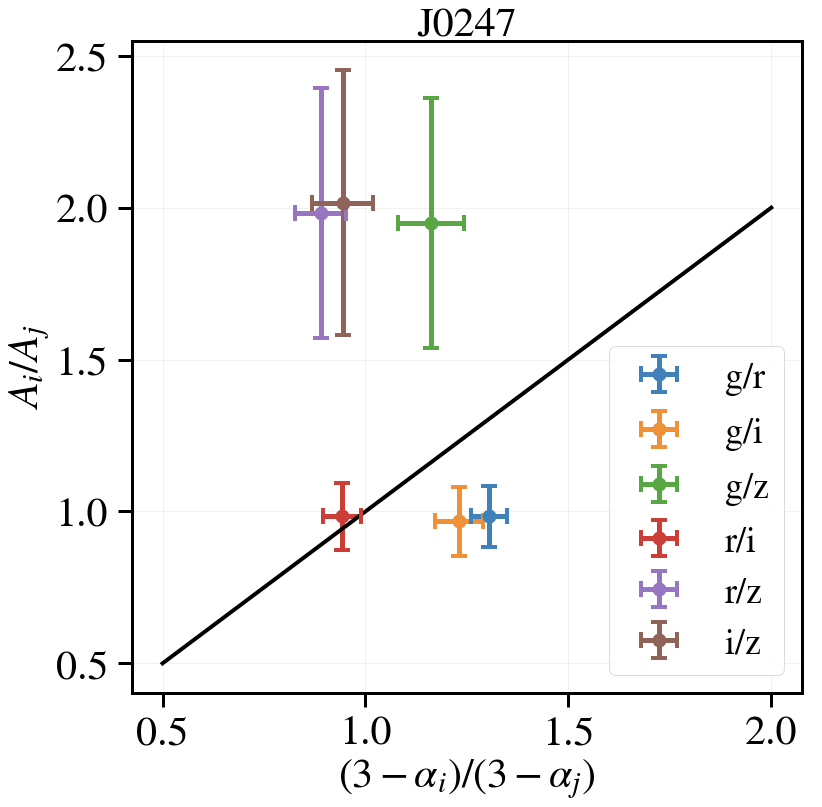}
\includegraphics[width=0.23\textwidth]{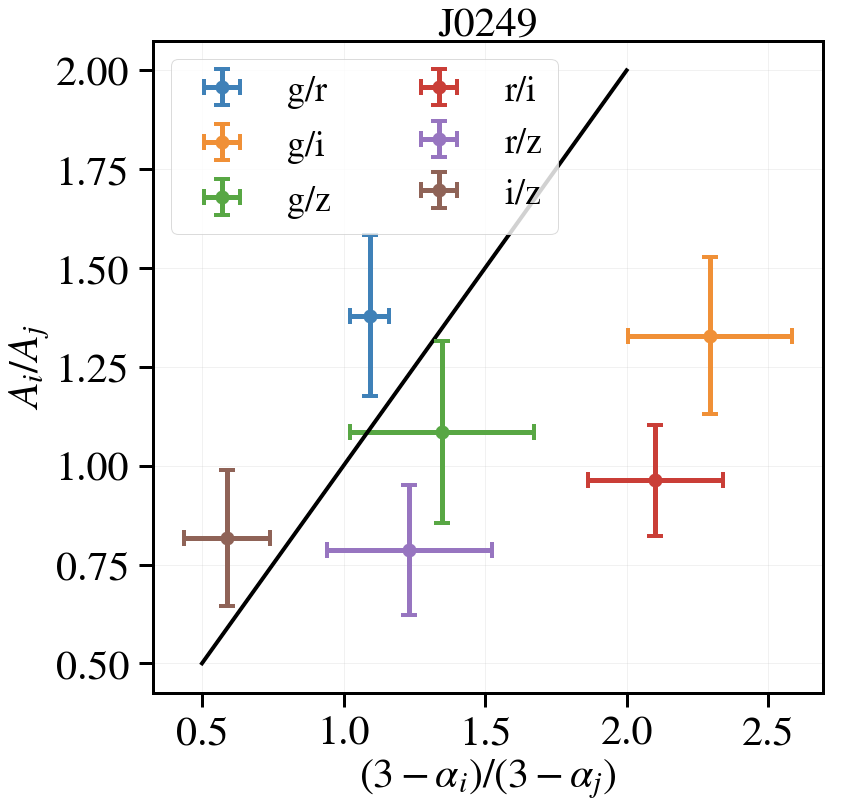}
\includegraphics[width=0.23\textwidth]{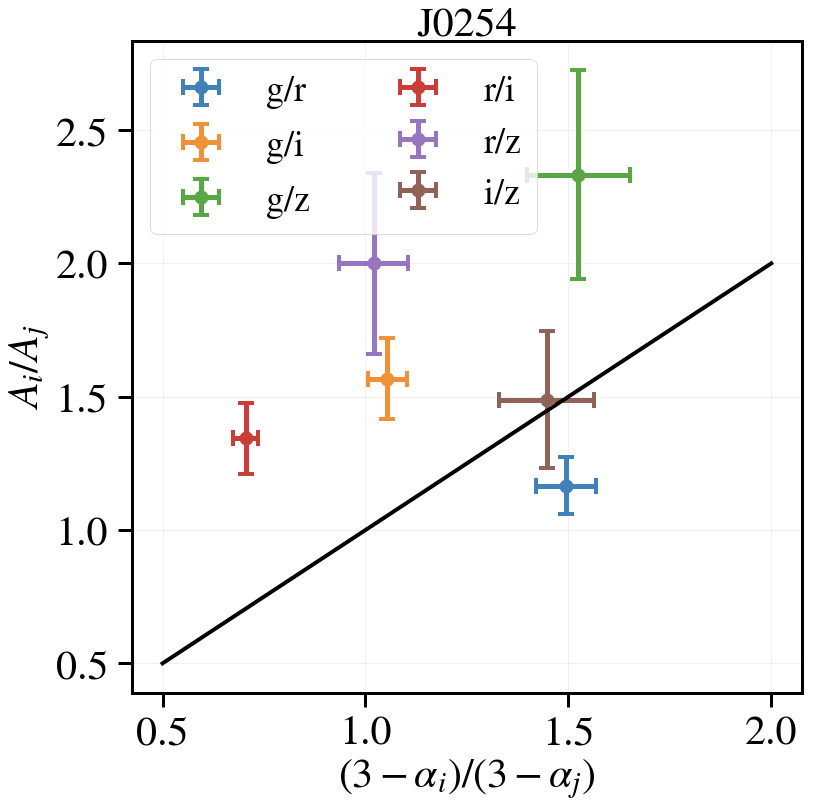}
\caption{Observed frequency-dependent variability amplitude ratios compared with the expected values from the Doppler boost model. Different colors represent different combinations of band pairs. The black lines represent the 1 to 1 relation. Error bars denote 1$\sigma$ uncertainties. See \S \ref{sec:doppler} for details.}
\label{fig:doppler}
\end{figure}

\autoref{tab:parameters} lists the multi-band spectral indices $\alpha_{\nu}$ from fitting a broken power-law model to the SDSS continuum spectrum and the variability amplitudes from the best-fit sinusoidal model. Although $\alpha_{\nu}$ could be itself variable due to variability, we assume that it is roughly constant in optical.  \autoref{fig:doppler} shows the observed multi-band variability amplitude ratios compared to the theoretical values expected from Doppler boost. In three of the four candidates (J0247, J0249 and J0254), the observed values show deviations from the expected values at the $\gtrsim2\sigma$ level in at least two band pairs with $p$-values of 0.189, 0.013 and 0.113. In J0246, the observed values are generally consistent with those expected from Doppler boost. 

\begin{figure*}
\centering
\includegraphics[width=0.49\textwidth]{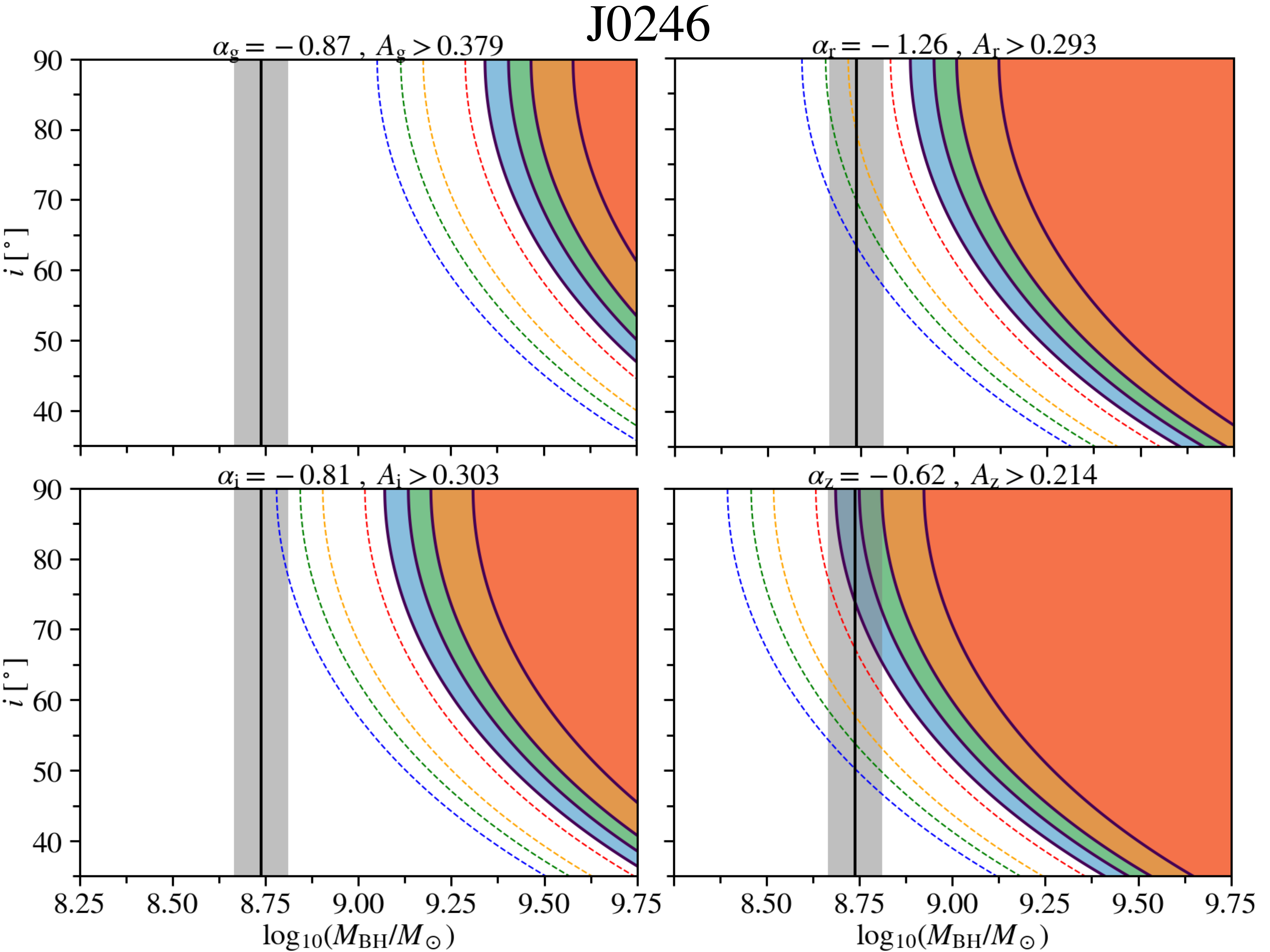}
\includegraphics[width=0.49\textwidth]{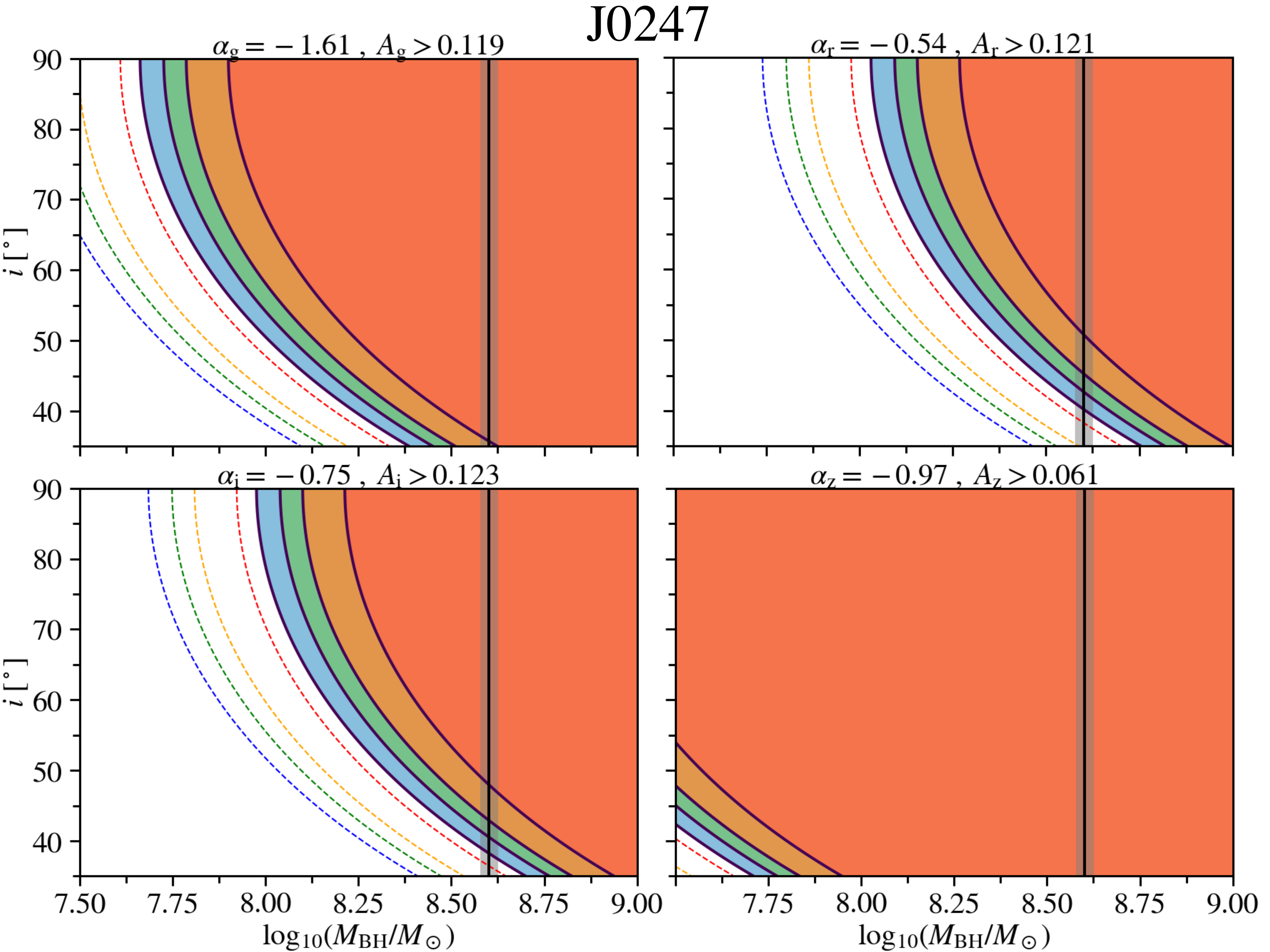}
\includegraphics[width=0.49\textwidth]{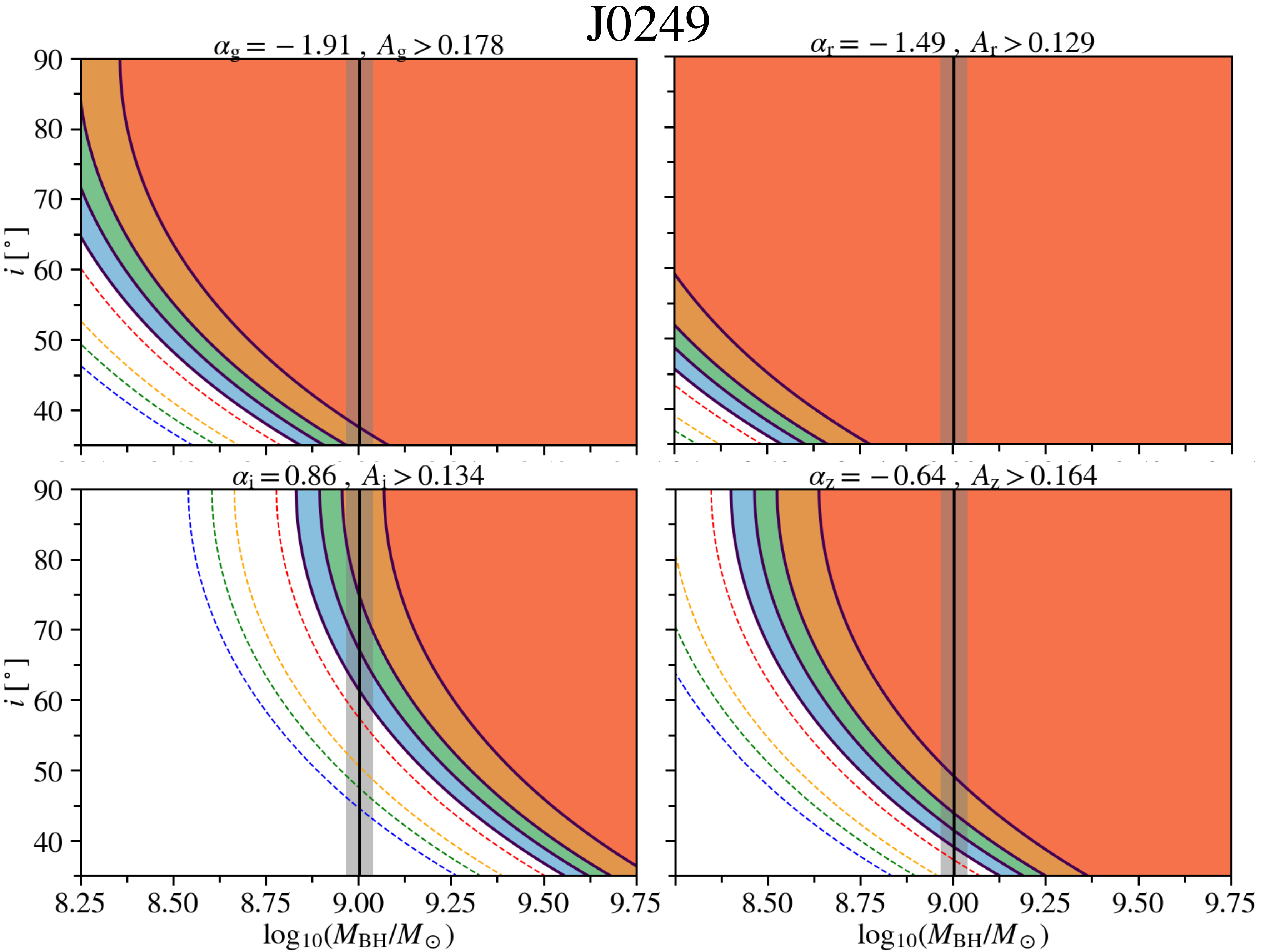}
\includegraphics[width=0.49\textwidth]{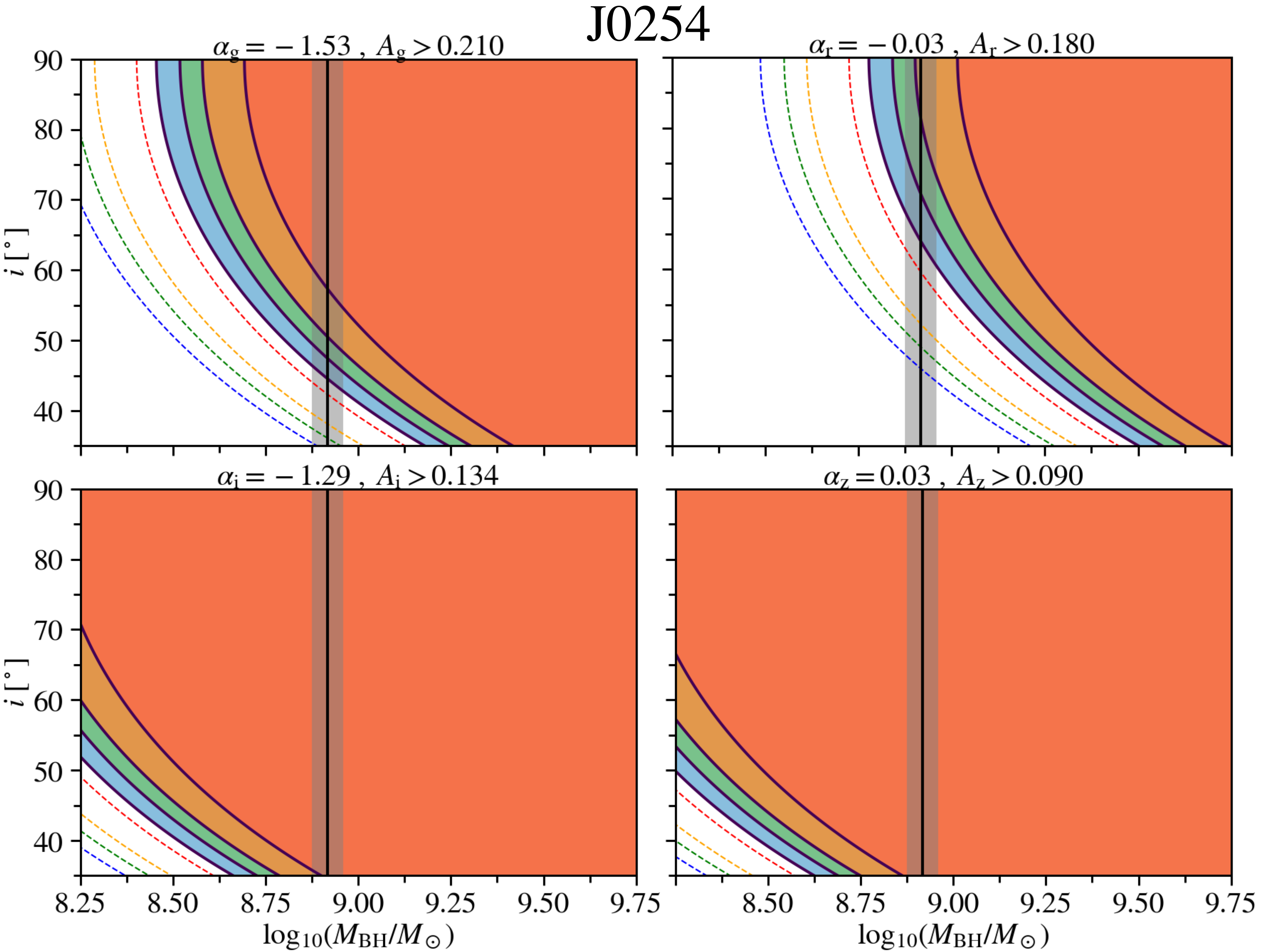}
\caption{Doppler boost model parameter space estimates for the four candidate periodic quasars. The four sub-panels for each quasar represent $griz$ bands. The dashed (shaded) contours denote $f_2$ = 1.0 ($f_2$ = 0.8), where $f_2$ is defined as the fraction of the total emission from the secondary black hole. Different colors represents different mass ratios with $q$=0.0, 0.05, 0.1, and 0.2 for blue, green, orange, and red, respectively. The black vertical line with grey shaded region shows the virial mass estimated from SDSS spectra (\autoref{fig:spectra}) and its 1$\sigma$ statistical error. The orbital inclination angle $i$=90$^{\circ}$ for an edge-on view and $i$=0$^{\circ}$ for a face-on view.}
\label{fig:doppler_pars}
\end{figure*}

\autoref{fig:doppler_pars} shows the allowed parameter space for the Doppler boost model to explain the observed multi-band variability amplitudes (\autoref{tab:parameters}). The parameters considered are the total black hole mass $M_{{\rm BH}}$, binary orbital inclination angle $i$, the fraction of the total emission from the secondary black hole $f_2$, and the mass ratio $q$. Our other fiducial parameters are the rest-frame orbital period $P_{{\rm orb}}$ = $P_{{\rm GLS}}/(1+z)$ and $\alpha_{\nu}$ in $griz$ bands (\autoref{tab:parameters}). 

For J0246, while the observed frequency-dependent variability amplitudes are generally consistent with the expected values under the Doppler boost hypothesis (\autoref{fig:doppler}), there is almost no allowed parameter space unless the following conditions are all met: 1. The virial black mass is significantly underestimated, even after considering a 0.4-dex systematic error \citep[e.g.,][]{Shen13_BHmass}, 2. A high fraction (e.g. $>80\%$) of the total emission is from the secondary black hole fueled by its mini accretion disk, and 3. The binary system is close to being an edge-on view. For the other three quasars, the observed multi-band variability amplitudes are generally consistent with the expected values under the Doppler boost hypothesis, although there is a stringent requirement on the allowed parameter space.

\citet{Liao2020} have tested the Doppler boost hypothesis for J0252. They suggest that the Doppler boost model is disfavored for J0252 based on both the frequency-dependent variability amplitude ratios and the multi-band variability amplitudes. 

In summary, there is evidence against the Doppler boost hypothesis in all five periodic candidates in our parent sample based on the observed frequency-dependent variability amplitude ratios (\autoref{fig:doppler}) and/or the multi-band variability amplitudes (\autoref{fig:doppler_pars}).


\section{Discussion}\label{sec:discussion}

\subsection{Comparison with Previous Systematic Searches of Candidate Periodic Quasars}\label{subsec:compare_previous_obs}

Previous optical light curve studies have found $\sim150$ candidate periodic quasars \citep{valtonen08,Graham2015a,Charisi2016,Zheng2016,Liutt2019,Li2019}. We compare our results with those from systematic searches based on large time-domain surveys. 

\citet{Graham2015a} have found 111 candidate periodic quasars out of 243,500 spectroscopically confirmed quasars in 33,000 deg$^2$ from the CRTS. They have adopted wavelet and ACF methods for the periodicity detection. They have requested $>$1.5 cycles to be covered by a total time baseline of nine years. Although \citet{Graham2015} had tailored simulation with 1000 mock DRW curves for the source PG 1302--102, due to the large size of the parent quasar sample, \citet{Graham2015a} have only produced one realization of simulated light curves for the whole sample to quantify the statistical significance of the periodicity detection. A major difference between our current work and \citet{Graham2015a} is that we have used a large set of simulated light curves with tailored variability parameters for each individual quasar to quantify the significance of periodicity detection. 



\citet{Zheng2016} have found one candidate periodic quasar out of a sample of 347 spectroscopically confirmed quasars in the SDSS-S82 based on light curves from the CRTS. They have adopted the GLSdeDRW and ACF methods for the period detection. While based on the same set of light curves from CRTS, the candidate identified by \citet{Zheng2016} was not selected by \citet{Graham2015a} as a periodic quasar. This suggests that different periodicity detection methods and different approaches to quantify the significance of any periodicity could lead to contrasting results based on the same data.

\citet{Charisi2016} have discovered 33 candidate periodic quasars out of 35,383 spectroscopically confirmed quasars in $\sim$3,000 deg$^2$ from the PTF. They have adopted the LS periodogram for periodicity detection. They have requested $>$1.5 cycles to be covered by a total time baseline of $\sim$4 years. As noted by \citet{Vaughan2016}, many candidate periodic quasars of \citet{Charisi2016} have small (i.e., yearly) periods that are subject to aliasing effects caused by the low cadence and seasonal gaps in the light curves. Unlike \citet{Charisi2016}, we have removed any periodicity detection with periods $<500$ days to minimize such artifacts.

\citet{Liutt2019} have initially identified 26 candidate periodic quasars from $\sim$9,000 photometrically selected quasars in $\sim$50 deg$^2$ from the PS1 Medium Deep survey. They have requested $>$1.5 cycles to be covered by a total time baseline of $\sim$4 years. Their re-analysis of the initial 26 candidates based on follow-up observations with extended time baselines have shown only one significant candidate periodic quasar. Instead of continued monitoring only for the initial periodic candidates selected based on the shorter baseline light curves, a more proper way should be to follow-up the entire parent sample for extended time baselines, because the initial periodic candidates could have all been false positives to begin with. In addition, only by following up the full parent sample can one be able to recover false negatives which could have been missed in the original selection. Finally, the quoted total numbers of cycles in the \citet{Liutt2019} re-analysis seem to be large at face value because the quoted ``periods'' were determined from the shorter baseline light curves, producing a misleadingly large number of ``cycles'' covered. A more proper reference of the number of cycles covered should have been based on the periods determined using the extended light curves.


Our detection rate is five out of 625, or ${\sim}0.8^{+0.5}_{-0.3}$\% assuming 1$\sigma$ Poisson error. This is $\sim$4--80 times higher than those found in previous studies at face value \citep[$\sim$0.01--0.2\%;][]{Graham2015a,Charisi2016,Zheng2016,Liutt2019}. An important factor that may explain the apparent difference in the detection rates found in our work compared to previous results is that our sample is probing a different quasar population than those studied in previous work.

\begin{figure}
\centering
\includegraphics[width=\columnwidth]{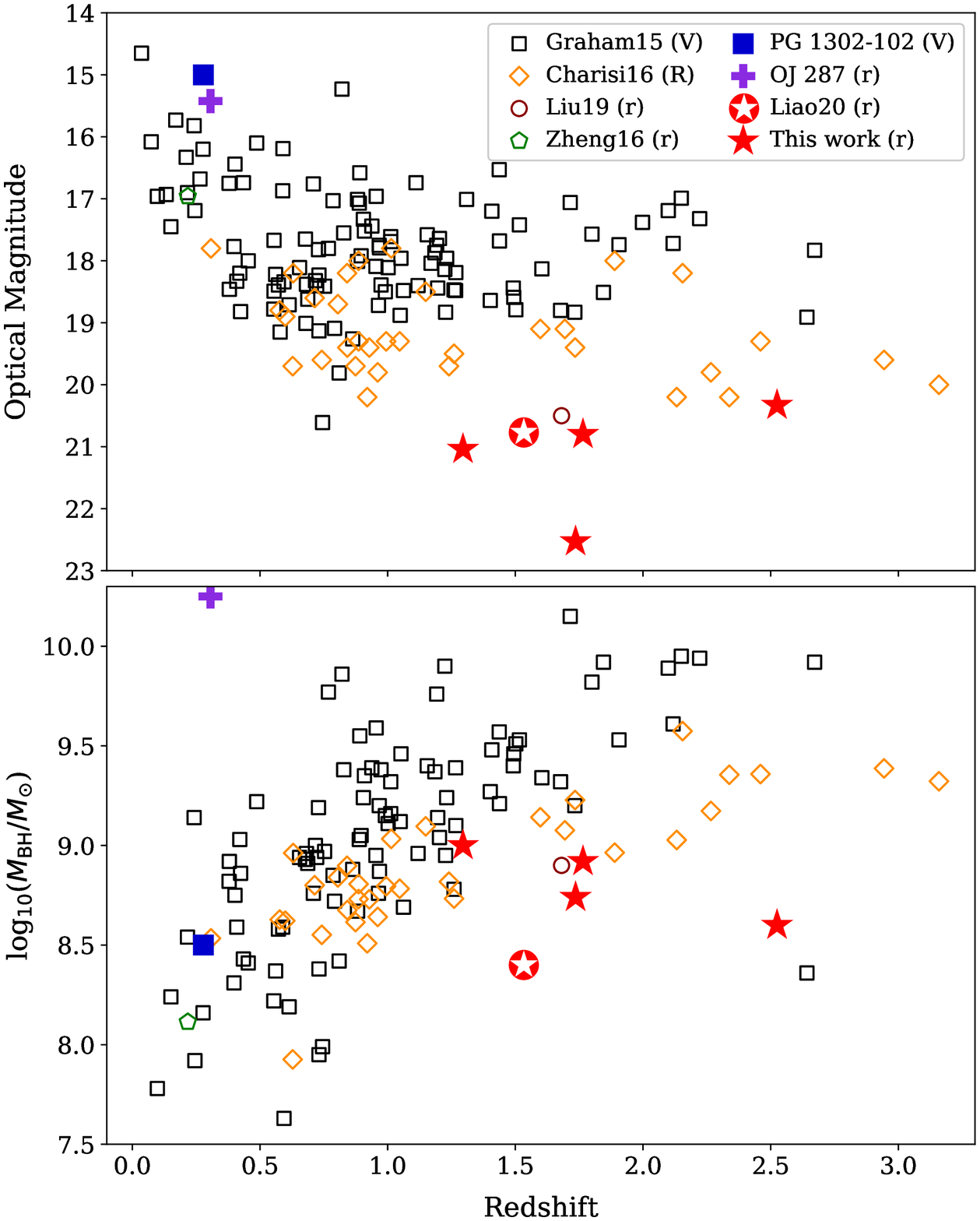}
\caption{\label{fig:mag_MBH_z} 
\textit{Top:} Redshift vs. optical magnitude of the four periodically variable quasars compared to those found by previous works \citep{Graham2015,Graham2015a,Charisi2016,Zheng2016,Liutt2019,Liao2020}. Our candidate periodic quasars are significantly fainter than those found from previous shallower surveys in larger areas. \textit{Bottom:} Redshift vs. BH mass estimate. Our candidate periodic quasars are systematically less massive (with $M_{{\rm BH}}{\sim}10^{8.4}$--$10^{9.0}M_{\odot}$) than those found by previous work at similar redshifts (with $M_{{\rm BH}}{\sim}10^9$--$10^{10}M_{\odot}$). At similar BH masses (with $M_{{\rm BH}}{\sim}10^{8.4}$--$10^{9.0}M_{\odot}$), our candidate periodic quasars are at larger redshifts ($z{\sim}2$) than those found by previous work ($z{\sim}0.7$).}
\end{figure}

\autoref{fig:mag_MBH_z} shows the redshift vs. magnitude of our candidate periodic quasars compared to those found in previous work. It demonstrates that our candidates are significantly fainter than those found from previous shallower surveys in larger areas.  \autoref{fig:mag_MBH_z} also shows the redshift vs. BH mass estimates of our candidate periodic quasars compared to those found in previous work. The BH masses of our candidate periodic quasars are systematically smaller than those found by previous work at similar redshift. At similar BH masses (with $M_{{\rm BH}}{\sim}10^{8.4}$--$10^{9.0}M_{\odot}$), our candidate periodic quasars are at larger redshifts ($z{\sim}2$) than those found by previous work ($z{\sim}0.7$). 

One possible explanation for the significantly higher detection rate of candidate periodic quasars in our sample -- assuming they are caused by BSBHs -- is that these less massive BHs (with $M_{{\rm BH}}{\sim}10^{8.4}$--$10^{9.0}M_{\odot}$) are still in the process of merging at redshift $z\sim$2, whereas the most massive BHs (with $M_{{\rm BH}}{\sim}10^9$--$10^{10}M_{\odot}$) probed by previous studies have already gone through their major merger process by $z\sim$2 due to cosmic ``downsizing'' in the evolution of SMBHs \citep[e.g.,][]{shen09}. At similar BH masses, the significantly higher detection fraction of periodic quasars found in our sample may be interpreted as the redshift evolution of the fraction of BSBHs, i.e., the binary fraction is larger at higher redshifts at a fixed BH mass consistent with theoretical expectations \citep[e.g.,][]{Kelley2019}.

Another possible reason to explain the apparent discrepancy is the differences in the quality of the light curve data of these various studies. While our parent sample size is smaller, which means that the statistical error of our detection rate is larger, the systematic uncertainty of our estimate is likely to be smaller, given that our data have longer time baselines and higher sensitivities. Higher data quality is more helpful not only for rejecting false positives but also for recovering false negatives. 

Finally, we also have relatively more stringent selection criteria by requesting at least three cycles to be spanned by the observations (\autoref{tab:compare}) and by using complementary selection criteria (\autoref{tab:number_candidates}), although this would have made our detection rate lower not higher than other estimates with everything else being equal.

In summary, we conclude that the differences in the quasar populations being probed and the quality of the light curve data are likely to be the dominating factors to explain the different detection rates of candidate periodic quasars in our sample compared to previous estimates.

\subsection{Comparison with Theoretical Predictions}\label{subsec:compare_theory}

Combining the results of \citet{Liao2020} and this paper, we have detected five candidate periodic quasars from a parent sample of 625 in a 4.6 deg$^2$ field. This corresponds to a detection rate of ${\sim}8^{+5}_{-3}$ per 10$^{3}$ quasars or ${\sim}1.1^{+0.7}_{-0.5}$ per deg$^2$. As discussed in \S \ref{subsec:alternative}, there are alternative explanations for the observed periodicity and so our detection rate is an upper limit of the rate of BSBHs.  Nevertheless, for the simplicity of discussion, we compare our observed detection rate to theoretical predictions of BSBHs assuming all candidate periodic quasars in our sample were caused by BSBHs.

\citet{haiman09} have considered circumbinary accretion and disk-binary interactions to model the orbital decay of BSBHs. According to Figure 8 of \citet{haiman09} assuming the $i$-band limit of $\sim$23.0 mag reached by our parent sample, we estimate $\lesssim$100 BSBHs in 2.7$\times$10$^2$ deg$^2$ with an observed period of $t_{var}$ = 60 weeks, which corresponds to $\lesssim$2 BSBHs in the 4.6 deg$^2$ field. Due to the limitation in the light curve time baseline and cadence, our detection window is between 500 days (to minimize aliasing effects) and $\sim$ 6 years (i.e., from the $>$3 cycles criterion), which is $\sim$1--5 times that of the assumed 60 week variability period ($t_{var}$) . Our observed detection rate is $\sim$2 times higher than but is broadly consistent with the theoretical rate predicted by \citet{haiman09} given uncertainties.


\citet{Volonteri2009} have studied the cosmic evolution of BSBHs using a Monte Carlo merger tree method and have traced the growth and dynamical history of BSBHs from high redshift. These authors have estimated 5--10 sub-pc BSBHs at $z<0.7$ in a sample of $10^4$ quasars. The rate may be a factor of 5--10 larger at $z>1$. However, the orbital timescale being considered is much longer than the periodicity that our search focuses since they consider the sub-pc population instead of the milli-pc population. 

\citet{Graham2015a} also provided the expected binary number of $\sim$450 in a survey sky coverage of 2$\pi$ ster with a detectable range of rest-frame orbital periods from 20--300 weeks, a limiting magnitude of V$\sim$20 and a redshift range of 0--4.5; it corresponds to an expected rate of 0.1 for 4.6 deg$^2$. Given our limiting r-band magnitude of 23 is three times deeper, the expected number should be in the same order of magnitude as our detection rate of 5 in 4.6 deg$^2$.

\citet{Kelley2019} have used a combination of cosmological simulations, semi-analytical binary merger models, and observed AGN properties to calculate the expected detection rate of periodic quasars. They have predicted 20 (from Doppler boost) and 100 (from hydrodynamic variability) BSBHs to be identified after 5 yr of LSST. Assuming the $r$-band limit of $\sim$23.5 mag reached in our parent sample (\autoref{fig:redshift_mag}), we estimate $\simeq$80 BSBHs in the full sky of $\sim$30,000 deg$^2$ with observed periods between 0.5 and 5.0 yr according to Figure 7 of \citet{Kelley2019}. The predicted detection rate is  ${\sim}2.6\times10^{-3}$ per deg$^2$, which is $\sim400$ times smaller than our detection rate. 
However, this is not a fair comparison because the numbers quoted above as predicted by \citet{Kelley2019} are appropriate for calculating the cumulative detection rate for an all-sky survey. Since we adopt a deep survey over a small area, our sample is dominated by the more common fainter quasars at high redshift ($z\sim2$). As a result, we should compare with the differential detection rate (which is a function of redshift and BH mass) appropriate for the quasar populations (i.e., less massive ones at high redshift) in our sample. This difference between the cumulative and differential detection rates is the same reason which likely explains the apparent discrepancy between our detection rate and those from previous shallower surveys over larger areas as discussed above (\S \ref{subsec:compare_previous_obs}).

In addition, as cautioned by \citet{Kelley2019}, the theoretical rate is still subject to unconstrained model assumptions. For example, the merger timescale as a function of binary separation is highly uncertain due to the lack of a self-consistent treatment of the accretion and dynamical evolution of BSBHs \citep[e.g.,][]{Dotti2012} in a cosmological context. It is also noted that our periodicity detection might not be complete for the periodic quasars with small amplitudes or close to the survey depth. However, to fully address this question, it requires detailed simulation and is beyond the scope of the current paper.


Finally, \citet{Kelley2019} have estimated that 1/6 of the periodic quasars from BSBHs would be caused by Doppler boost whereas 5/6 would be due to circumbinary accretion variability. \citet{Liao2020} has shown for the case of J0252 that circumbinary accretion variability is strongly preferred over Doppler boost based on both BIC analysis of the light curve model fitting and the frequency-dependent variability amplitude tests. For the other four candidate periodic quasars, our BIC analyses of the light curve model fittings do not show a strong preference for circumbinary accretion variability over Doppler boost. For our four candidates, the Doppler boost hypothesis is disfavored based on the frequency-dependent variability amplitude ratios and/or the multi-band variability amplitudes (\S \ref{sec:doppler}). While we still cannot rule out Doppler boost as a possible origin for the observed periodicity in the five candidate periodic quasars, we do find tentative evidence that it is not the dominant mechanism for driving the periodicity seen in the majority of our sample, which is consistent with the prediction of \citet{Kelley2019}.


\subsection{Implications for the Final-Parsec Problem and Gravitational Waves}\label{subsec:gw}

\autoref{tab:info} lists the estimated binary separations assuming that the binary is in a circular orbit and that the orbital period equals the observed light curve periodicity as expected for circumbinary accretion disk variability with a mass ratio of $q=0.11$ \citep{Farris2014}. If these candidate periodic quasars were indeed caused by BSBHs, then the inferred millipc-scale binary separations imply that the systems would have already passed the ``final-parsec'' barrier \citep{begelman80}.

The GW strain amplitude of a circular binary in the quadrupole approximation can be expressed as
\begin{equation}
    h_0 = \frac{4G}{c^2}\frac{M_c}{D_L}\left(\frac{G}{c^3}2\pi f_{\rm orb} M_c\right)^{2/3} , 
\end{equation}
where the chirp mass is
\begin{equation}
    M_c = \left[\frac{q}{(q+1)^2}\right]^{3/5} M_{{\rm BH}},
\end{equation}
$G$ is the gravitational constant, $D_L$ is the luminosity distance, and $f_{\rm orb}$ is the orbital frequency. 
Using our mass estimation for the four quasars ( \autoref{tab:info}), the inferred GW strain amplitudes are $\sim$ 3.7$\times$10$^{-18}$, 1.5$\times$10$^{-18}$, 1.5$\times$10$^{-17}$, 6.7$\times$10$^{-17}$ for J0246, J0247, J0249 and J0254 assuming $q=0.11$. The GW strain amplitudes of our candidate periodic quasars are more than two order of magnitude lower than the PTA upper limits \citep{Zhu2014,Arzoumanian2018a}, making them undetectable by current PTAs as individual sources. If $q=1$, the inferred amplitudes would be three times larger, but would still be far from the current PTA limits. This is unsurprising because our sample selected in a small field only probes the more common and less massive black holes that are not massive enough to be detectable by PTAs. While our BSBH candidates cannot be detected individually, recent PTA upper limits on the stochastic background have been used to constrain the fraction of BSBH candidates from variability surveys \citep[e.g.,][]{Sesana2018,Holgado2018}. As PTA sensitivities improve, the space density of the most massive milli-pc BSBH candidates will be further constrained.

Assuming our candidate periodic quasars are caused by BSBHs in the GW-driven regime, the coalescence timescale for a circular binary is given by
\begin{equation}
t_{\rm gw} = \frac{5}{256} \left(\frac{G M_c}{c^3}\right)^{-5/3} (2 \pi f_{\rm orb})^{-8/3} \ .
\end{equation}
\autoref{tab:parameters} lists the estimated GW inspiral times $t_{{\rm gw}}$ for our candidate periodic quasars, which are much shorter than the Hubble time. This implies that the candidate binaries are efficiently emitting GWs and will merge within the age of the universe, even if environmental effects are neglected \citep[e.g.,][]{holgado_gravitational_2019}. The inferred $t_{{\rm gw}}$ (${\gtrsim}10^4$ yr) is ${\sim}10^3$ times shorter than estimates of quasar lifetimes \citep{Martini2001,Yu2002}. This is reasonable considering that the suggested probability of observing them based on the timescale ratio is not too smaller than their detection rate.  

\begin{figure}
\centering
\includegraphics[width=\columnwidth]{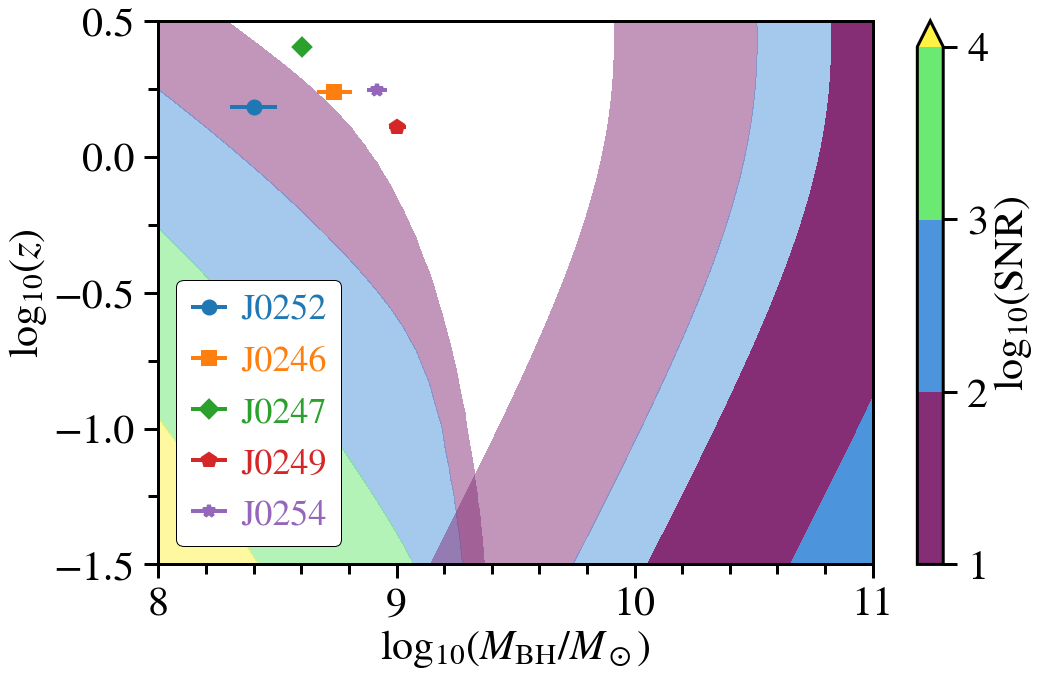}
\caption{\label{fig:sensitivity} 
Signal-to-noise ratio (SNR) contours for GW sources assuming the candidate periodic quasars are caused by merging BSBHs. The transparent contours are for 2030s detectors like LISA and an SKA-like facility. The opaque contours are for NANOGrav, which is currently operating. Except for J0252 (which is the least massive), our candidate periodic quasars fall in the sensitivity gap between LISA and future SKA-like PTA.
}
\end{figure}

\autoref{fig:sensitivity} shows signal-to-noise ratio (SNR) for future GW detectors as a functions of black hole mass and redshift. We calculate the SNRs for LISA, the NANOGrav 11-yr limits, and a SKA-like facility using the gwent package\footnote{\url{https://gwent.readthedocs.io/}} (Kaiser and McWilliams, in prep). Our candidate periodic quasars would mostly fall in the sensitivity gap between LISA and future PTAs. LISA would only be sensitive to a J0252-like merger which is the least massive system among our five candidate periodic quasars. 

\subsection{Spectral Energy Distribution}\label{subsec:sed}

Given the total BH masses of the four quasars (${\sim}$10$^{8.6}$ M$_\odot$--10$^{9.0}$M$_\odot$; \autoref{tab:info}) and assuming the observed light curve periods of a few years represent the orbital periods (\autoref{tab:parameters}; \S \ref{subsec:gw}), the candidate BSBHs are expected at pre-decoupling \citep{Kocsis2011,Tanaka2012,Sesana2012}, where circumbinary accretion disks should be common. One theoretical prediction of circumbinary accretion disk models is abnormalities (e.g., flux deficits or cutoffs) in the IR-optical-UV SEDs due to a central cavity opened by the binaries and/or minidisks around both BHs \citep[e.g.,][]{Roedig2014,Foord2017,Tang2018}.

\begin{figure*}
\centering
\includegraphics[width=0.49\textwidth]{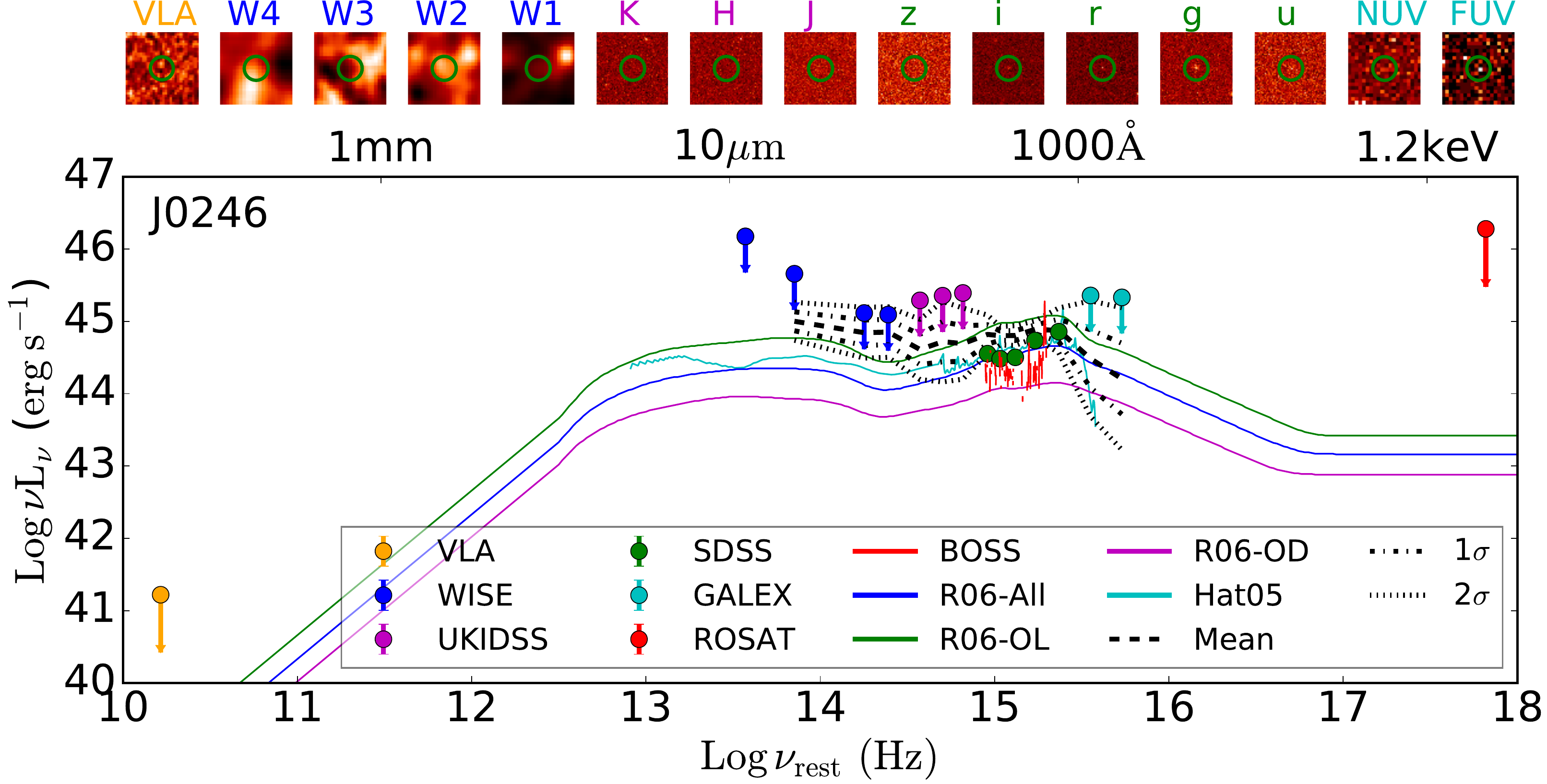}
\includegraphics[width=0.49\textwidth]{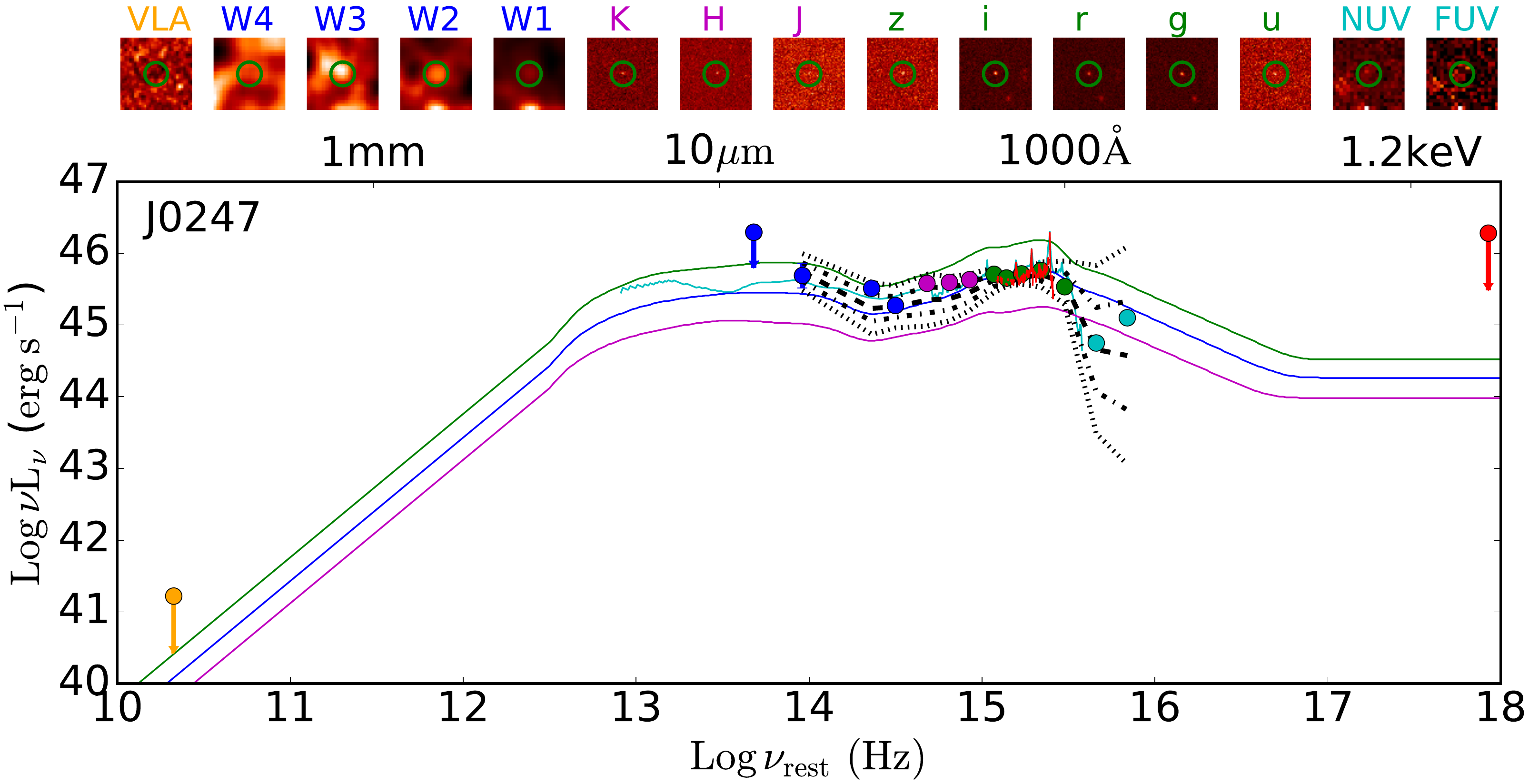}
\includegraphics[width=0.49\textwidth]{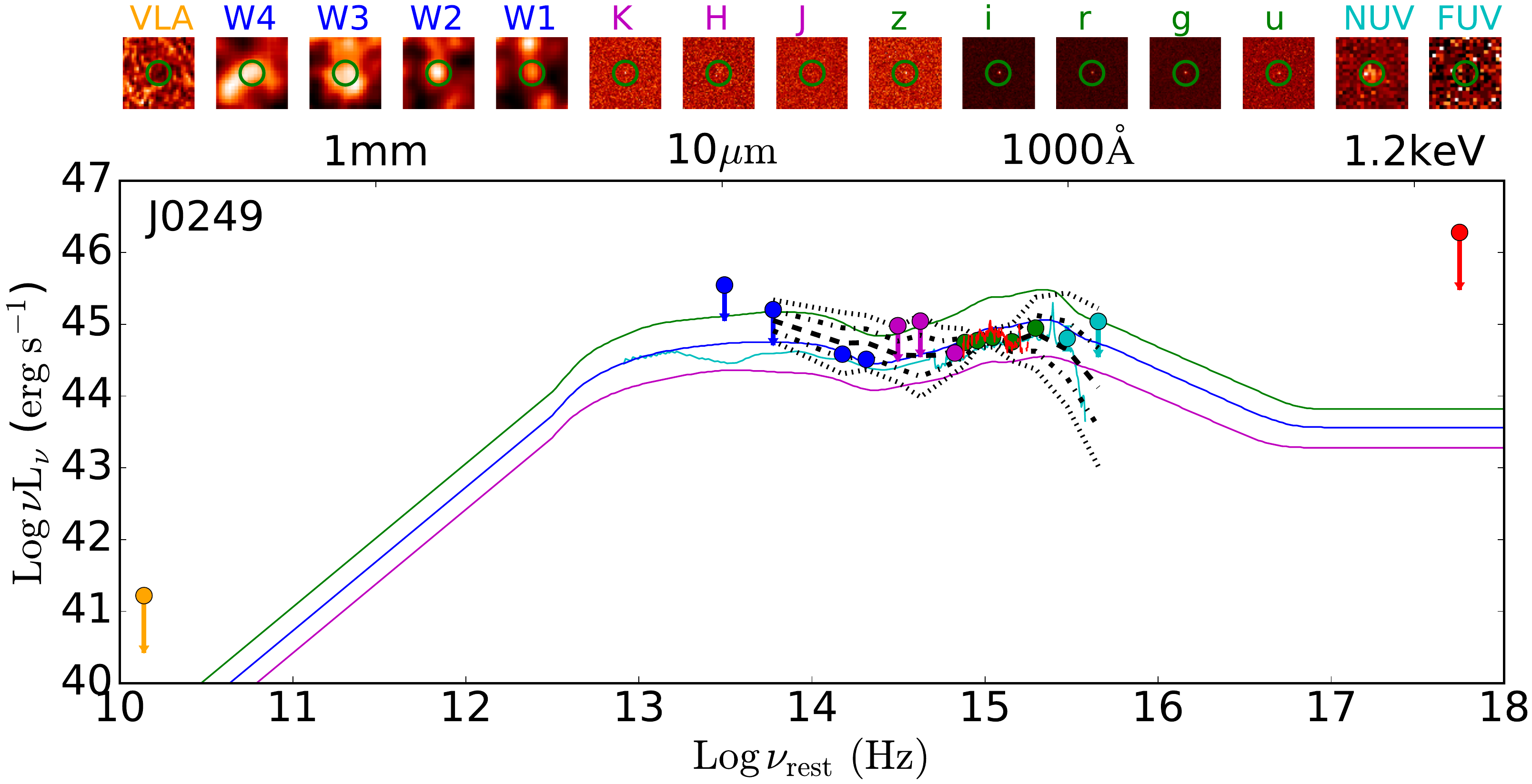}
\includegraphics[width=0.49\textwidth]{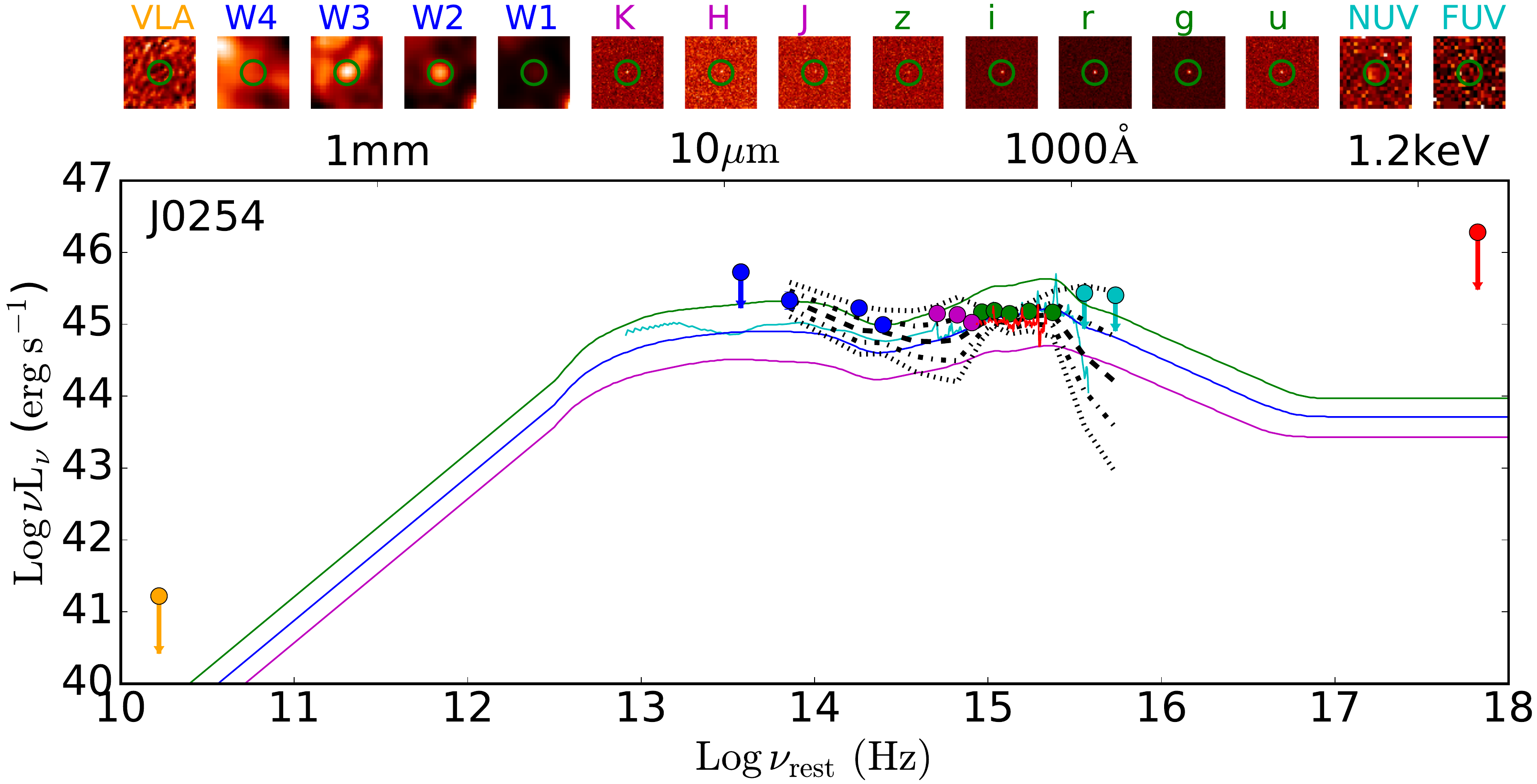}
\caption{
Multi-wavelength SEDs of the four candidate periodic quasars. Shown for comparison are the median SEDs for a control quasar sample (black curves) matched in redshift and luminosity from optically selected quasars \citep{Richards2006}, the mean SED of the whole sample in \citet{Richards2006} (``R06-All'' for all quasars, ``R06-OL'' for optically luminous quasars, and ``R06-OD'' for optically dim quasars), and the mean SED from \citet{Hatziminaoglou2005} (Hat05). Upper limits are 3$\sigma$. Shown on top of each sub-panel are the multi-wavelength postage stamp images with an FOV of 30$''\times$30$''$ each. The green circles are 10$''$ in diameter centered at the quasar's optical coordinates.}
\label{fig:sed}
\end{figure*}

\autoref{fig:sed} shows the multi-wavelength SEDs of the four candidate periodic quasars. The SED data include radio flux density upper limits from the VLA sky survey \citep[VLASS,][]{VLASS}, mid-infrared photometry from the Wide-field Infrared Survey Explorer \citep[WISE,][]{Wright2010}, near-infrared photometry from the UKIRT Infrared Deep Sky Survey \citep[UKIDSS,][]{Lawrence2007}, optical photometry and optical spectra from the SDSS \citep{York2000}, UV photometry from the Galaxy Evolution Explorer \citep[GALEX,][]{Martin2005}, and X-ray upper limits from ROSAT \citep{voges00}. None of the four quasars has radio/X-ray detection from VLASS/ROSAT. Only two of the four have UV detections from GALEX. 


\autoref{fig:sed} also shows the mean SEDs of control samples of ordinary quasars that are drawn to match the redshifts and luminosities of our candidate periodic quasars. All four candidate periodic quasars show similar SEDs (within 3$\sigma$) to those of the control quasars. \citet{Guo2020sed} have studied the SEDs of the periodic quasar candidates discovered by \citet{Graham2015} and \citet{Charisi2016}. They have found that the SEDs of periodic quasar candidates from the CRTS and PTF studies are consistent with those of the control quasars. Similarly, none of our periodic quasar candidates shows any SED peculiarity compared to the control quasars, although the existing SED data may not be good enough to detect such abnormal features in the SEDs.


\subsection{Alternative Interpretations}\label{subsec:alternative}


We discuss alternative interpretations that are not caused by BSBHs to explain the candidate periodic light curves.


\subsubsection{Precessing Radio Jets}\label{subsubsec:jet}

The optical flux of the quasars could be the sum of thermal emission from the accretion disk and the non-thermal emission from the Doppler boosting radio jet \citep{Rieger2006}. If the accretion disk is misaligned with the plane of the symmetry, the non-thermal jet precession could lead to periodic variation in the optical flux. 

Unlike previously known periodic candidates OJ287 and PG1302 \citep{valtonen08,Graham2015a}, our candidate periodic quasars are not known to be blazars. None of the four quasars was detected by VLASS \citep{VLASS}. The implied 3-$\sigma$ flux density upper limit is $f^{obs}_{3GHz}$ $<$0.36 mJy. To estimate the upper limit of the radio loudness parameter \citep{Kellermann1989} $R\equiv f_{6cm}/f_{2500}$, we calculate the rest-frame flux density at 5 GHz (6 cm corresponding to 5 GHz) assuming the radio flux follows a power law $f_{\nu}\propto v^{\alpha}$. \autoref{tab:info} lists the inferred radio loudness upper limits for the four quasars. Combining with the $f_{2500}$ measurements from the optical spectra, the inferred $R$ are $<$272, $<$62, $<$85, and $<$52 for J0246, J0247, J0249, and J254 assuming a spectral index of $\alpha=-0.5$ \citep{Jiang2007}. While the existing radio loudness upper limits cannot exclude the possibility of the four quasars being radio loud ($R>$10), they do suggest that the optical emission is not dominated from a radio jet (i.e., $R>$100 \citep{Chiaberge2011}) except for J0246. The precessing radio jet scenario is therefore unlikely for three of four candidate periodic quasars, although future deeper radio imaging is still needed to completely rule out such a scenario.

\subsubsection{Tilted Accretion Disks}

Besides radio jets, precession of a tilted (warped) accretion disk could also produce periodic flux variation by obscuring the continuum-emitting region. \citet{Tremaine2014} have estimated the timescale of $1.3\times10^5$yr for the warped disk around a $10^8 M_{\odot}$ BH to damp out and align with the BH axis. The damping timescale is much shorter than the typical AGN timescale. This suggests that the warped disk behavior is unlikely to be seen in our quasar sample. Assuming the warped disk scenario and without any external torque, self-gravity is expected to play an important role in the disk precession \citep{Tremaine2014}. \citet{Ulybay-Siddiki09} have estimated that the precessing rate of self-gravitating warped disk around a BH is given by,
\begin{equation}
    \dot{\phi} \sim C\frac{M_{{\rm disk}}}{M_{{\rm BH}}}\bigg(\frac{GM_{{\rm BH}}}{r_w^3}\bigg)^{1/2},
\end{equation}
where $M_{{\rm BH}}$ is the black hole mass, $M_{{\rm disk}}$ is the disk mass, $r_w$ is the radius of the warped disk, and $C$ is the constant of order unity depending on the disk mass configuration. \citet{Tremaine2014} have derived the warped radius $r_w$ of 290 gravitational radius $r_g\equiv GM_{BH}/c^2$ for a $10^8 M_{\odot}$ BH using fiducial parameters. Assuming $M_{disk}/M_{BH} = 0.01$, the precessing period is ${\sim}50$ yr. While the period is an order of magnitude larger than the periods we have found in our search, it is still broadly consistent considering model uncertainties. However, the amount of obscured continuum emission required would be too large to explain the observed variability amplitudes seen in our candidate periodic quasars (\autoref{tab:parameters}).

\subsubsection{Quasi-periodic Oscillations}\label{subsubsec:qpo}

Quasi-periodic oscillations (QPOs) are commonly seen in X-ray stellar binaries due to strong resonances in the accretion flow. Low-frequency QPO is thought to be related to the Lense-Thirring precession \citep{Bardeen1975,Ingram2009}. Assuming period of QPO is approximately proportional to black hole mass, \citet{King2013} discovered the first AGN scaled-up version of QPO with period of 120$-$150 days in the radio light curve at 15 GHz. With the similar scaling relation, our candidates with BH masses of ${\sim}10^{8.4}$--$10^{9.0}M_{\odot}$ give QPOs with the period of few hundreds to thousands days, which are in the same order of the rest-frame periods we found. However, QPOs seen in the X-ray binaries show drifts in amplitude, period, and/or phase \citep{vanderKlis1989}. Future continuous monitoring is needed to distinguish BSBHs from QPOs.

\section{Summary and Future Work}\label{sec:sum}

We have presented a systematic search for periodically variable quasars in a sample of 625 spectroscopically confirmed quasars at a median redshift of $z{\sim}1.8$ in the overlapping regions between DES-SN S1 and S2, and SDSS-S82 fields (4.6 deg$^2$). Our search is based on a unique 20-year long time baseline by combining new DES-SN multi-band ($griz$) light curves with archival SDSS-S82 observations. Our light curve data also have better sensitivities than those from previous surveys over larger areas. The deep imaging has allowed us to search for periodic light curves in less luminous quasars (down to $r{\sim}$23.5 mag; \autoref{fig:redshift_mag}) powered by less massive SMBHs (with $M_{{\rm BH}}\gtrsim10^{8.5}M_{\odot}$) at high redshift for the first time. We summarize our main findings in the following. 

\begin{enumerate}
    \item We have discovered five candidate periodic quasars in a parent sample of 625 quasars. The most significant candidate (J0252) has been presented by \citet{Liao2020}. Here we have presented the other four candidates (\S \ref{subsec:result_periodicity}). Our selection criteria are: 1. Significant (at the $>$99.74\% level) periodicity in the GLS periodogram in at least two bands with at least three cycles covered by the total time baseline (to reduce false positives due to stochastic variability) and a period larger than 500 days (to minimize artifacts due to aliasing). 2. Consistent periodicity in ACF. and 3. the variability amplitude of the periodic component is larger than that in the residual from fitting the light curves with a sinusoidal model. 

    \item We have found that the fraction of periodic quasars in our parent sample is five out of 625 which corresponds to ${\sim}0.8^{+0.5}_{-0.3}$\% assuming 1$\sigma$ Poisson error, or ${\sim}1.1^{+0.7}_{-0.5}$ per deg$^2$. We have estimated that up to two of the five candidates may be false positives due to red noise accounting for the look-elsewhere effect (\S \ref{subsec:result_significance}). 

    \item We have estimated the virial BH masses of the four candidate periodic quasars based on single-epoch estimator using spectral modeling of the SDSS spectra (\S \ref{sec:bhmass}). Their estimated BH masses are ${\sim}10^{8.6}$--$10^{9.0}M_{\odot}$ (\autoref{tab:info}). Together with J0252 \citep{Liao2020}, our candidate periodic quasars are systematically less massive (with $M_{{\rm BH}}{\sim}10^{8.4}$--$10^{9.0}M_{\odot}$) than those found by previous works at similar redshifts (with $M_{{\rm BH}}{\sim}10^9$--$10^{10}M_{\odot}$; \autoref{fig:mag_MBH_z}). At similar BH masses (with $M_{{\rm BH}}{\sim}10^{8.4}$--$10^{9.0}M_{\odot}$), our candidate periodic quasars are at higher redshifts ($z{\sim}2$) than those found by previous works ($z{\sim}0.7$; \autoref{fig:mag_MBH_z}).
    
    \item Our periodicity detection rate is $\sim$4--80 times higher than those found in previous searches using shallower surveys over larger areas \citep[0.1--2 per 10$^3$ quasars;][]{Graham2015,Charisi2016,Zheng2016,Liutt2019}. The apparent discrepancy is likely caused by differences in the quasar populations being probed (\autoref{fig:mag_MBH_z}) and the quality of the light curve data from various surveys (\S \ref{subsec:compare_previous_obs}).
    
    \item Assuming the periodicity is caused by BSBHs, our periodicity detection rate is consistent with the theoretical prediction of \citet{haiman09} within uncertainties. While our periodicity detection rate is $\sim400$ times higher than the prediction of \citet{Kelley2019} taken at face value, the large apparent discrepancy is likely caused by the difference in the cumulative and differential detection rates (\S \ref{subsec:compare_theory}), since the binary fraction may vary as a function of redshift and BH mass. 
    
    
    \item As an alternative test, we have used a maximum likelihood approach to assess if an additional periodic signal is needed to explain the light curves on top of a stochastic background (\S \ref{subsec:lc_model}). We have considered two physically different models for the periodic signal, i.e., Doppler and hydrodynamic variability. In one of the four candidate periodic quasars (J0246), both periodic models being considered are strongly preferred over a stochastic variability model. In the other three candidates, however, there is no preference for the two periodic models over a stochastic variability model, although we cannot rule out the possibility that the data may still prefer other forms of periodic models over stochastic variability.
    
    
    \item We have compared the observed frequency-dependent variability amplitude ratios  and the multi-band variability amplitudes with those predicted from the Doppler boost hypothesis \citep[e.g.,][]{DOrazio2015}. We have found evidence against the Doppler boost model in all four candidate periodic quasars (\S \ref{sec:doppler}). For J0246, while the observed frequency-dependent variability amplitude ratios are generally consistent with Doppler boost (\autoref{fig:doppler}), the observed variability amplitudes are too large to be explained with typical model parameters (\autoref{fig:doppler_pars}). For the other three candidate periodic quasars, the observed frequency-dependent variability amplitude ratios show ${\gtrsim}2\sigma$ deviations in at least two band pairs (\autoref{fig:doppler}). While the observed variability amplitudes can generally be explained by Doppler boost, the required parameters need some level of fine tuning (\autoref{fig:doppler_pars}).
    
    
    \item Assuming the optical flux periodicity is caused by BSBHs, we have estimated the binary separations assuming the orbital periods equal the flux periods and circular orbits with a mass ratio of 0.11 (\autoref{tab:info}). If the periodicity were indeed caused by BSBHs, the inferred millipc-scale separations would imply that the final-parsec barrier has been overcome in these quasars. The inferred GW inspiral times are $\gtrsim10^4$ yr (\autoref{tab:info}) for the four candidates presented here and is $\sim10^5$ yr for J0252 \citep{Liao2020}. We have discussed the prospect of detecting the potential GWs from these candidate periodic quasars with current and future low-frequency GW experiments (\S \ref{subsec:gw}). Except for J0252 (the least massive),  our candidate periodic quasars would fall in the sensitivity gap between LISA and future PTAs as individual sources (\autoref{fig:sensitivity}), which is unsurprising given their that they are too massive for LISA and yet not massive enough and too far away for PTAs (\autoref{fig:mag_MBH_z}).
    
    
    \item We have examined the multi-wavelength SEDs of our candidate periodic quasars to search for evidence for possible abnormalities as predicted by circumbinary accretion disk models (\S \ref{subsec:sed}). All four quasars show SEDs that are consistent with those of control quasars matched in redshift and luminosity (\autoref{fig:sed}).
    
    \item We have discussed alternative interpretations for the optical flux periodicity that are not caused by BSBHs (\S \ref{subsec:alternative}). While the existing data generally disfavor the precessing-radio-jet and tilted-accretion-disk scenarios, they are consistent with optical QPOs for the four candidate periodic quasars.
    
\end{enumerate}


Continued photometric monitoring is needed to further assess the robustness of the periodicity discovered here. While the existing time baseline spans $\sim$4--6 cycles (we have required $>$three cycles by selection) of the periodicity in our candidates (\autoref{tab:parameters}), there are seasonal gaps and coverage gaps between surveys (\autoref{fig:lc_and_periodogram}). In addition, the existing SDSS imaging was generally too noisy to distinguish between different light curve models, while the high-quality DES imaging alone was not long enough to cover enough number of cycles. Continued monitoring with our ongoing program with DECam on the Blanco 4m telescope and the LSST in future with the Vera Rubin Observatory will improve both the light curve coverage and the data quality. Continuous monitoring will be carried out for the full parent sample. This is important for both rejecting false positives and to recovering false negatives. Future monitoring will also help distinguish periodic light curves caused by BSBHs and optical QPOs (\S \ref{subsubsec:qpo}).

In future work, we will enlarge our parent sample size by studying photometrically selected quasars (combining color and variability selections) in the DES-SN and other deep fields with long time baselines by combining archival observations. This will allow us to select more candidate periodic quasars and to reduce the statistical error of their detection rate (\S \ref{subsec:compare_previous_obs}). 

Future more sensitive UV and X-ray observations are needed to tighten the constraints on the potential SED abnormalities to test theoretical predictions from circumbinary accretion disk models \citep{Milosavljevic2005,Roedig2012,Foord2017,Tang2018}. Deep radio imaging is also required to further constrain the processing-radio-jet scenario (\S \ref{subsubsec:jet}).

More realistic theoretical light curve models from self-consistent 3D circumbinary accretion disk simulations are needed to compare with future observations with better data quality. Finally, alternative methods for robust periodicity detection \citep[e.g.,][]{Zhu2020} can independently test the significance of the candidate periodic quasars proposed in this work. 

\section*{Acknowledgements}
We thank the anonymous referee for comments and suggestions. X.L. thanks A. Barth, B. Fields, C. Gammie, Z. Haiman, D. Lai, A. Loeb, D. Orazio, S. Tremaine, and X.-J. Zhu for discussions. Y.-C.C. and X.L. acknowledge a Center for Advanced Study Beckman fellowship and support from the University of Illinois campus research board. A.M.H. is supported by the DOE NNSA Stewardship Science Graduate Fellowship under grant number DE-NA0003864. Y.S. acknowledges support from the Alfred P. Sloan Foundation and NSF grant AST-1715579. This work makes extensive use of SDSS-I/II and SDSS-III/IV data (http://www.sdss.org/). 

Funding for the DES Projects has been provided by the U.S. Department of Energy, the U.S. National Science Foundation, the Ministry of Science and Education of Spain, 
the Science and Technology Facilities Council of the United Kingdom, the Higher Education Funding Council for England, the National Center for Supercomputing 
Applications at the University of Illinois at Urbana-Champaign, the Kavli Institute of Cosmological Physics at the University of Chicago, 
the Center for Cosmology and Astro-Particle Physics at the Ohio State University,
the Mitchell Institute for Fundamental Physics and Astronomy at Texas A\&M University, Financiadora de Estudos e Projetos, 
Funda{\c c}{\~a}o Carlos Chagas Filho de Amparo {\`a} Pesquisa do Estado do Rio de Janeiro, Conselho Nacional de Desenvolvimento Cient{\'i}fico e Tecnol{\'o}gico and 
the Minist{\'e}rio da Ci{\^e}ncia, Tecnologia e Inova{\c c}{\~a}o, the Deutsche Forschungsgemeinschaft and the Collaborating Institutions in the Dark Energy Survey.

The Collaborating Institutions are Argonne National Laboratory, the University of California at Santa Cruz, the University of Cambridge, Centro de Investigaciones Energ{\'e}ticas, 
Medioambientales y Tecnol{\'o}gicas-Madrid, the University of Chicago, University College London, the DES-Brazil Consortium, the University of Edinburgh, 
the Eidgen{\"o}ssische Technische Hochschule (ETH) Z{\"u}rich, 
Fermi National Accelerator Laboratory, the University of Illinois at Urbana-Champaign, the Institut de Ci{\`e}ncies de l'Espai (IEEC/CSIC), 
the Institut de F{\'i}sica d'Altes Energies, Lawrence Berkeley National Laboratory, the Ludwig-Maximilians Universit{\"a}t M{\"u}nchen and the associated Excellence Cluster Universe, 
the University of Michigan, NFS's NOIRLab, the University of Nottingham, The Ohio State University, the University of Pennsylvania, the University of Portsmouth, 
SLAC National Accelerator Laboratory, Stanford University, the University of Sussex, Texas A\&M University, and the OzDES Membership Consortium.

Based in part on observations at Cerro Tololo Inter-American Observatory at NSF’s NOIRLab (NOIRLab Prop. ID 2012B-0001; PI: J. Frieman), which is managed by the Association of Universities for Research in Astronomy (AURA) under a cooperative agreement with the National Science Foundation.

The DES data management system is supported by the National Science Foundation under Grant Numbers AST-1138766 and AST-1536171.
The DES participants from Spanish institutions are partially supported by MICINN under grants ESP2017-89838, PGC2018-094773, PGC2018-102021, SEV-2016-0588, SEV-2016-0597, and MDM-2015-0509, some of which include ERDF funds from the European Union. IFAE is partially funded by the CERCA program of the Generalitat de Catalunya.
Research leading to these results has received funding from the European Research
Council under the European Union's Seventh Framework Program (FP7/2007-2013) including ERC grant agreements 240672, 291329, and 306478.
We acknowledge support from the Brazilian Instituto Nacional de Ci\^encia
e Tecnologia (INCT) do e-Universo (CNPq grant 465376/2014-2).

This manuscript has been authored by Fermi Research Alliance, LLC under Contract No. DE-AC02-07CH11359 with the U.S. Department of Energy, Office of Science, Office of High Energy Physics.

We are grateful for the extraordinary contributions of our CTIO colleagues and the DECam Construction, Commissioning and Science Verification
teams in achieving the excellent instrument and telescope conditions that have made this work possible.  The success of this project also 
relies critically on the expertise and dedication of the DES Data Management group.



\section*{Data availability}

The data underlying this article will be shared on reasonable request to the corresponding author. The light curves of the four candidates (J0246, J0247, J0249 and J0254) are available in the article's online supplementary material.



\bibliographystyle{mnras}
\bibliography{ref,binaryrefs} 




\appendix


\bsp	
\label{lastpage}
\end{document}